\documentclass[twocolumn]{aastex62}
\usepackage{amsmath}
\usepackage{empheq}
\usepackage{mathrsfs}
\usepackage{textcomp}
\usepackage{enumitem}   
\usepackage{gensymb}
\usepackage{hyperref}
\usepackage{graphicx}
\usepackage[caption=false]{subfig}
\usepackage{multirow}
\usepackage{longtable}
\hypersetup{colorlinks, linkcolor={blue}, citecolor={blue}, urlcolor={blue}}  
\definecolor{DarkOrange}{RGB}{204, 85, 0}
\definecolor{LincolnGreen}{RGB}{17, 102, 0}
\def\ion#1#2{#1$\;${\footnotesize\rm{#2}}\relax}

\begin{document}

\title{ZTF Early Observations of Type Ia Supernovae I: \\ Properties of the 2018 Sample}


\author[0000-0001-6747-8509]{Yuhan Yao}
\affiliation{Cahill Center for Astrophysics, California Institute of Technology, MC 249-17, 1200 E California Boulevard, Pasadena, CA, 91125, USA}
\email{yyao@astro.caltech.edu}

\author[0000-0001-9515-478X]{Adam A. Miller}
\affiliation{Center for Interdisciplinary Exploration and Research in Astrophysics (CIERA) and Department of Physics and Astronomy, Northwestern University, 2145 Sheridan Road, Evanston, IL 60208, USA}
\affiliation{The Adler Planetarium, Chicago, IL 60605, USA}

\author[0000-0001-5390-8563]{S. R. Kulkarni}
\affiliation{Cahill Center for Astrophysics, California Institute of Technology, MC 249-17, 1200 E California Boulevard, Pasadena, CA, 91125, USA}

\author[0000-0002-8255-5127]{Mattia Bulla}
\affiliation{The Oskar Klein Centre, Department of Physics, Stockholm University, AlbaNova, SE-106 91 Stockholm, Sweden}

\author[0000-0002-8532-9395]{Frank J. Masci}
\affiliation{IPAC, California Institute of Technology, 1200 E. California Blvd, Pasadena, CA 91125, USA}

\author[0000-0003-3461-8661]{Daniel A. Goldstein}
\affiliation{Cahill Center for Astrophysics, California Institute of Technology, MC 249-17, 1200 E California Boulevard, Pasadena, CA, 91125, USA}

\author[0000-0002-4163-4996]{Ariel Goobar}
\affiliation{The Oskar Klein Centre, Department of Physics, Stockholm University, AlbaNova, SE-106 91 Stockholm, Sweden}

\author[0000-0002-3389-0586]{Peter Nugent}
\affiliation{Computational Cosmology Center, Lawrence Berkeley National Laboratory, 1 Cyclotron Road, Berkeley, CA 94720, USA}
\affiliation{Department of Astronomy, University of California, Berkeley, CA 94720-3411, USA}

\author{Alison Dugas}
\affiliation{Cahill Center for Astrophysics, California Institute of Technology, MC 249-17, 1200 E California Boulevard, Pasadena, CA, 91125, USA}

\author[0000-0003-0901-1606]{Nadia Blagorodnova}
\affiliation{Department of Astrophysics/IMAPP, Radboud University, Nijmegen, The Netherlands}

\author[0000-0002-0466-1119]{James D. Neill}
\affiliation{Cahill Center for Astrophysics, California Institute of Technology, MC 249-17, 1200 E California Boulevard, Pasadena, CA, 91125, USA}

\author{Michael Rigault}
\affiliation{Universit\'e Clermont Auvergne, CNRS/IN2P3, Laboratoire de Physique de Clermont, F-63000 Clermont-Ferrand, France}

\author{Jesper Sollerman}
\affiliation{The Oskar Klein Centre, Department of Astronomy, Stockholm University, AlbaNova, 10691 Stockholm, Sweden}

\author{J. Nordin}
\affiliation{Institute of Physics, Humboldt-Universit{\"a}t zu Berlin, Newtonstr. 15, 12489 Berlin, Germany}

\author[0000-0001-8018-5348]{Eric C. Bellm}
\affiliation{DIRAC Institute, Department of Astronomy, University of Washington, 3910 15th Avenue NE, Seattle, WA 98195, USA} 

\author[0000-0003-1673-970X]{S. Bradley Cenko}
\affiliation{Astrophysics Science Division, NASA Goddard Space Flight Center, 8800 Greenbelt Road, Greenbelt, MD 20771, USA}
\affiliation{Joint Space-Science Institute, University of Maryland, College Park, MD 20742, USA}

\author[0000-0002-8989-0542]{Kishalay De}
\affiliation{Cahill Center for Astrophysics, California Institute of Technology, MC 249-17, 1200 E California Boulevard, Pasadena, CA, 91125, USA}

\author{Suhail Dhawan}
\affiliation{The Oskar Klein Centre, Department of Physics, Stockholm University, AlbaNova, SE-106 91 Stockholm, Sweden}

\author[0000-0002-9435-2167]{Ulrich Feindt}
\affiliation{The Oskar Klein Centre, Department of Physics, Stockholm University, AlbaNova, SE-106 91 Stockholm, Sweden}  

\author[0000-0002-4223-103X]{C. Fremling}
\affiliation{Cahill Center for Astrophysics, California Institute of Technology, MC 249-17, 1200 E California Boulevard, Pasadena, CA, 91125, USA}

\author[0000-0002-1955-2230]{Pradip Gatkine}
\affiliation{Department of Astronomy, University of Maryland College Park, MD 20742, USA}

\author[0000-0002-3168-0139]{Matthew J. Graham}
\affiliation{Division of Physics, Mathematics, and Astronomy, California Institute of Technology, Pasadena, CA 91125, USA}

\author[0000-0002-9154-3136]{Melissa L. Graham}
\affiliation{University of Washington, Department of Astronomy Box 351580 Seattle WA 98195-1580, USA}

\author[0000-0002-9017-3567]{Anna Y. Q.~Ho}
\affiliation{Cahill Center for Astrophysics, California Institute of Technology, MC 249-17, 1200 E California Boulevard, Pasadena, CA, 91125, USA}

\author[0000-0002-9878-7889]{T.\ Hung}
\affiliation{Department of Astronomy and Astrophysics, University of California, Santa Cruz, CA 95064, USA}

\author[0000-0002-5619-4938]{Mansi M. Kasliwal}
\affiliation{Cahill Center for Astrophysics, California Institute of Technology, MC 249-17, 1200 E California Boulevard, Pasadena, CA, 91125, USA}

\author[0000-0002-6540-1484]{Thomas Kupfer}
\affiliation{Kavli Institute for Theoretical Physics, University of California Santa-Barbara, Santa Barbara, CA, USA}

\author[0000-0003-2451-5482]{Russ R. Laher}
\affiliation{IPAC, California Institute of Technology, 1200 E. California Blvd, Pasadena, CA 91125, USA}

\author[0000-0001-8472-1996]{Daniel A. Perley}
\affiliation{Astrophysics Research Institute, Liverpool John Moores University,\\ IC2, Liverpool Science Park, 146 Brownlow Hill, Liverpool L3 5RF, UK}

\author[0000-0001-7648-4142]{Ben Rusholme}
\affiliation{IPAC, California Institute of Technology, 1200 E. California Blvd, Pasadena, CA 91125, USA}

\author[0000-0003-4401-0430]{David L. Shupe}
\affiliation{IPAC, California Institute of Technology, 1200 E. California Blvd, Pasadena, CA 91125, USA}

\author[0000-0001-6753-1488]{Maayane T. Soumagnac}
\affiliation{Benoziyo Center for Astrophysics, Weizmann Institute of Science, Rehovot, Israel}

\author[0000-0002-5748-4558]{K. Taggart}
\affiliation{Astrophysics Research Institute, Liverpool John Moores University,\\ IC2, Liverpool Science Park, 146 Brownlow Hill, Liverpool L3 5RF, UK}

\author{Richard Walters}
\affiliation{Cahill Center for Astrophysics, California Institute of Technology, MC 249-17, 1200 E California Boulevard, Pasadena, CA, 91125, USA}
\affiliation{Caltech Optical Observatories, California Institute of Technology, MC 249-17, 1200 E California Boulevard, Pasadena, CA, 91125, USA}

\author[0000-0002-0786-7307]{Lin Yan}
\affiliation{Cahill Center for Astrophysics, California Institute of Technology, MC 249-17, 1200 E California Boulevard, Pasadena, CA, 91125, USA}

\begin{abstract}
Early-time observations of Type Ia supernovae (SNe Ia) are essential to constrain their progenitor properties. In this paper, we present high-quality light curves of 127 SNe Ia discovered by the Zwicky Transient Facility (ZTF) in 2018. We describe our method to perform forced point spread function (PSF) photometry, which can be applied to other types of extragalactic transients. With a planned cadence of six observations per night ($3g+3r$), all of the 127 SNe Ia are detected in both $g$ and $r$ band more than 10\,d (in the rest frame) prior to the epoch of $g$-band maximum light. The redshifts of these objects range from $z=0.0181$ to 0.165; the median redshift is 0.074. Among the 127 SNe, 50 are detected at least 14\,d prior to maximum light (in the rest frame), with a subset of 9 objects being detected more than 17\,d before $g$-band peak. This is the largest sample of young SNe Ia collected to date; it can be used to study the shape and color evolution of the rising light curves in unprecedented detail. We discuss six peculiar events in this sample, including one 02cx-like event ZTF18abclfee (SN\,2018crl), one Ia-CSM SN ZTF18aaykjei (SN\,2018cxk), and four objects with possible super-Chandrasekhar mass progenitors: ZTF18abhpgje (SN\,2018eul), ZTF18abdpvnd (SN\,2018dvf), ZTF18aawpcel (SN\,2018cir) and ZTF18abddmrf (SN\,2018dsx).
\end{abstract}

\section{Introduction}
Despite being used as standardizable candles to study cosmology, the origin of Type Ia supernovae (SNe Ia) is not settled \citep[see review by][]{Maoz2014}. Two major formation channels have been proposed: single degenerate (SD), where a carbon/oxygen white dwarf (WD) accretes matter from a non-degenerate star triggering an explosion near the Chandrasekhar mass ($M_{\rm ch}$, \citealt{Whelan1973}); and double degenerate (DD), in which the primary WD accretes material from (or merges with) another WD \citep{Woosley1994, Tutukov1996, Shen2015}.

Observations obtained in the hours to days after explosion (i.e., ``early-time'') provide a path towards diagnosing the various explosion mechanisms \citep{Maoz2014}. Early photometry can constrain the radii of the possible companion and the progenitor star \citep{Kasen2010, Nugent2011, Bloom2012, Goobar2014, Goobar2015}. The shape and duration of the rising light curves probe the radial distribution of radioactive $^{56}$Ni in the exploding core, as well as the existence of circumstellar material \citep{Piro2014, Dessart2014, Firth2015, Piro2016, Miller2018}. 

Simulations of the double detonation of a helium shell on the surface of a WD predict an unusually red excess well before peak luminosity \citep{Noebauer2017, Maeda2018, Polin2019}, which were observed in SN\,2016jhr \citep{Jiang2017} and SN\,2018byg \citep{De2019}. Should the SD channel hold, a collision between the SN ejecta and the stellar companion will give rise to strong ultraviolet (UV) emission at early times \citep{Kasen2010, Hayden2010a}. The detection of a declining UV pulse in the peculiar SN Ia iPTF\,14atg reasonably favors this scenario \citep{Cao2015}, although \citet{Kromer2016} argue that its spectral evolution is more consistent with a merger product. The power of well-sampled early-time photometry has also been demonstrated in single-object studies of normal SNe Ia \citep{Marion2016, Hosseinzadeh2017, Shappee2019, Dimitriadis2019}, where the clearly-resolved early bumps in their light curves pose challenges to simple explosion models.

Up to now there has not been a large ($>100$ objects) uniform dataset of SN Ia light curves with both multi-band photometry and dense early-time sampling. With the Zwicky Transient Facility (ZTF, \citealt{Bellm2019b, Graham2019}), we are undertaking a high-cadence survey with six epochs per night ($3g+3r$; \citealt{Bellm2019a}). This experiment is conducted over a large area of the sky ($\sim$2500\,deg$^2$) and thus enables large-number statistics. In this study, we focus on a special subset of SNe Ia that were discovered more than 10\,d prior to maximum light. Our large (127 objects), homogeneous sample of young SNe Ia was constructed within the first year of operations by ZTF.

As the first in a series of three papers, we present the light curves and sample properties of 127 SNe Ia. A detailed analysis of the early evolution of these SNe will be addressed in Miller et al.~(in prep) and Bulla et al.~(in prep). Throughout this paper, we assume a flat $\Lambda$CDM cosmology with $H_0= 73.24 \, \rm km \, s^{-1}\, Mpc^{-1}$ \citep{Riess2016} and $\Omega_m = 0.275$ \citep{Amanullah2010}.

\section{The ZTF 2018 High-Cadence SN Ia Sample} \label{sec:sample}
\begin{deluxetable*}{clc}[htp]
\tablecaption{Sample Selection Steps\label{tab:selection}}
\tablehead{
\colhead{Step}
& \colhead{\ \ \ \ \ Criteria \ \ \ \ \ }
&\colhead{Total \# SNe}
} 
\startdata
    1 & Spectroscopically classified SNe Ia observed by the ZTF partnership survey  & 336\\
    2 & Observed by the high-cadence fields & 247 \\
    3 & At least one detection on the Marshal light curve earlier than 5\,d prior to $t_{B, \rm max}$ & 191\\
    4 & Remove observations where the reference images are obtained after $t_{B,\rm max}-25$\,d & 154\\
    \hline
    \multicolumn{3}{c}{-- Extract foced-PSF photometry light curves -- }\\
    \hline
    5 & Before $t_{B, \rm max}$, the target must be detected in both $g$ and $r$ over at least five nights & 140\\
    6 & The first 3-$\sigma$ detection in both $g$ and $r$ must be earlier than $t_{B, \rm max}-10(1+z)$ & 129\\
    7 & In [$t_{B, \rm max}$, $t_{B, \rm max}+20(1+z)$], must be detected at least once in both $g$ and $r$  &127\\
\enddata
\end{deluxetable*}

\subsection{Observations} 
The ZTF camera is mounted on the 48-inch Samuel Oschin Telescope (P48) at Palomar Observatory (\citealt{Dekany2016}, Dekany et al. submitted). At a limiting magnitude of $r\sim 20.5$\,mag, three custom filters ($g_{\rm ZTF}$, $r_{\rm ZTF}$, and $i_{\rm ZTF}$; hereafter $g$, $r$, and $i$) are designed to maximize throughput by avoiding major skylines at Palomar \citep{Bellm2019b}. ZTF divides its observing time between public surveys (40\%), partnership surveys (40\%), and Caltech surveys (20\%). \citet{Bellm2019a} provide details of the ZTF surveys. In brief, 85\% of the public time was allocated to a ``Northern Sky Survey'' with a 3-day cadence in $g$ and $r$, and the remaining 15\% to a ``Galactic Plane Survey'' with two visits of the Galactic plane ($1\, g + 1\, r$) every night. The bulk of the partnership time (May--December) in 2018 was dedicated to two experiments, including an extragalactic high-cadence experiment covering $\sim$2500 deg$^2$ with six visits ($3 \, g+3\,r$) every night, and a lower-cadence, wide field $i$-band survey. The Caltech time was conducted in a one-day cadence in both $g$ and $r$ with a total footprint of $\sim$3000\,deg$^2$. Each night's schedule is arranged by the survey scheduler \citep{Bellm2019a} to optimize volumetric survey speed \citep{Bellm2016}. In this work, we only focus on the $g$- and $r$-band observations within the partnership high-cadence fields. 

The current ZTF alert distribution system (\citealt{Patterson2019}) generates a source packet once a transient\footnote{For the purpose of this paper, moving objects are ignored.} is detected. By definition, a ``detection'' means that the observed flux is 5 times larger than the flux uncertainty (see \citealt{Masci2019}, Section 6). For each transient the alert packet includes a rolling 30-day history of detections and non-detections.

Following the association of all alerts generated at the same position, the GROWTH ``Marshal'' \citep{Kasliwal2019} compiles a complete historical record of variability, which is further used to aggregate and visualize follow-up observations. 

\subsection{Initial Sample Selection} \label{subsec:initial sample selection}
The sample selection process is summarized in Table \ref{tab:selection}. In total, there were 336 SNe Ia classified in the partnership fields in 2018, 247 of which were observed as part of the high-cadence partnership survey.\footnote{A measurement of the detection efficiency and completeness of this sample is beyond the scope of this study and will be addressed in a future paper (Nordin et al., 2019, in prep).} For the sample of 247 SNe with high-cadence observations, we performed a preliminary fit to their light curves using the \texttt{SALT2} software package \citep{Guy2007} implemented in the \texttt{sncosmo} python package \citep{Barbary2016}\footnote{\url{https://sncosmo.readthedocs.io/en/v2.0.x/models.html}} to estimate the time of maximum light. The sample was further reduced to include only those sources with greater than one detection in either the $g$ or $r$ band obtained at least 5\,d before the \texttt{SALT2}-estimated time of $B$-band maximum, $t_{B, \mathrm{max}}$. This resulted in a selection of 191 SNe. 

Note that although the 191 SNe were all discovered in the partnership high cadence fields, some of them also have observations under the public or Caltech time. We retained those observations in the following analysis.

The ZTF Science Data System (ZSDS) constructed reference images for each field and filter by taking the stack-average of 15--40 historical images\footnote{The ZTF camera has 16 CCDs, with each CCD divided in four quadrants.} \citep{Masci2019}. Observations for each target could be covered by multiple fields. To ensure that the reference images do not contain contamination from SN flux, we need to treat each field (with specific CCD-quadrant therein) for a given filter separately. Hereafter we use ``fcqf\,ID'', defined by 
\begin{align}
  (\rm fcqf\,ID) =&(\rm field \, ID) \times 10000 + (\rm CCD \,ID) \times  100 \, +   \notag \\
 &(\rm quadrant \,ID) \times 10 + (\rm filter\, ID) \label{eq:fcqf}
\end{align}
as an identifier of the reference images.

We grouped observations by fcqf\,ID, and excluded those if the time of the latest exposure used to create the reference product was within 25\,d of $t_{B, \rm max}$ (in the SN rest frame). Since the typical rise time of SNe Ia is 17\,d \citep{Firth2015}, 25\,d is a conservative choice. For each target, we further required that the remaining number of observations in both $g$ and $r$ must be no less than 35. 154 SNe met this criterion. 

\section{Data Analysis} \label{sec:analysis}
\begin{figure}
    \centering
    \includegraphics[width=\columnwidth]{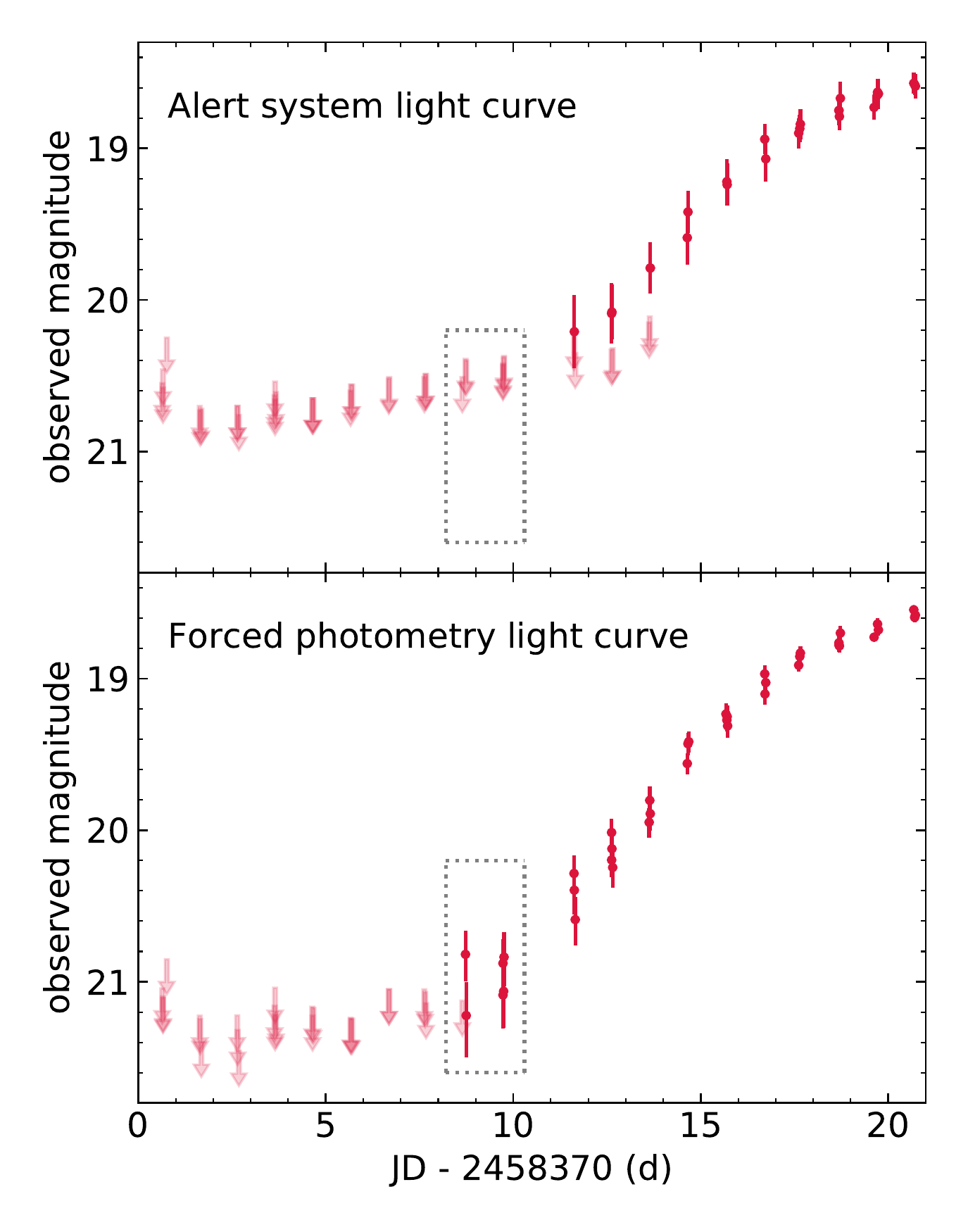}
    \caption{Upper panel: the $r$-band light curve of ZTF18abxxssh generated by the alert distribution system. Bottom panel: forced-PSF photometry light curve of the same object. The dotted box highlights additional early-time $r$-band detections recovered by forced photometry.  \label{fig:whyforcephot}}
\end{figure}

We perform ``forced'' PSF photometry to extract precise flux measurements of the SN in all ZTF images, including those that were obtained prior to explosion. Forced PSF light curves are obtained by measuring the PSF flux at the position of the SN in all epochs. Figure~\ref{fig:whyforcephot} demonstrates the difference between the light curves generated by the ZTF alert packets and forced-PSF photometry. The improvement is clear as highlighted by the dotted box, where the forced photometry recovers detections that are otherwise missed by the real-time pipeline. It is also the case that the forced photometry provides deeper pre-explosion upper limits. Thus, forced-PSF photometry can (i) provide sub-threshold flux measurements, (ii) reveal structure in the early-time light curves, and (iii) allow more stringent constraints to be placed on the epoch of explosion. 
\subsection{Astrometry}
The position of ZTF transients reported on the GROWTH Marshall is based on the intial detection of the source, which is often at low signal-to-noise ratio (SNR). As the first step of forced-PSF photometry, we need to determine the position of the transient more accurately. To this end, for each SN, we obtained the coordinates in all epochs where the SN is detected using \texttt{Kowalski},\footnote{\url{https://github.com/dmitryduev/kowalski}} a ZTF database system. The typical scatter in both R.A.\ and Decl.\ is $\sim$0.088 arcsec (0.09 pixel size). This uncertainty is small, and thus we do not incorporate it into the PSF modelling. We took the median R.A.\ and Decl.\ as the true position of each SN. 

\begin{figure*}[htb]
    \centering
    \includegraphics[width=\textwidth]{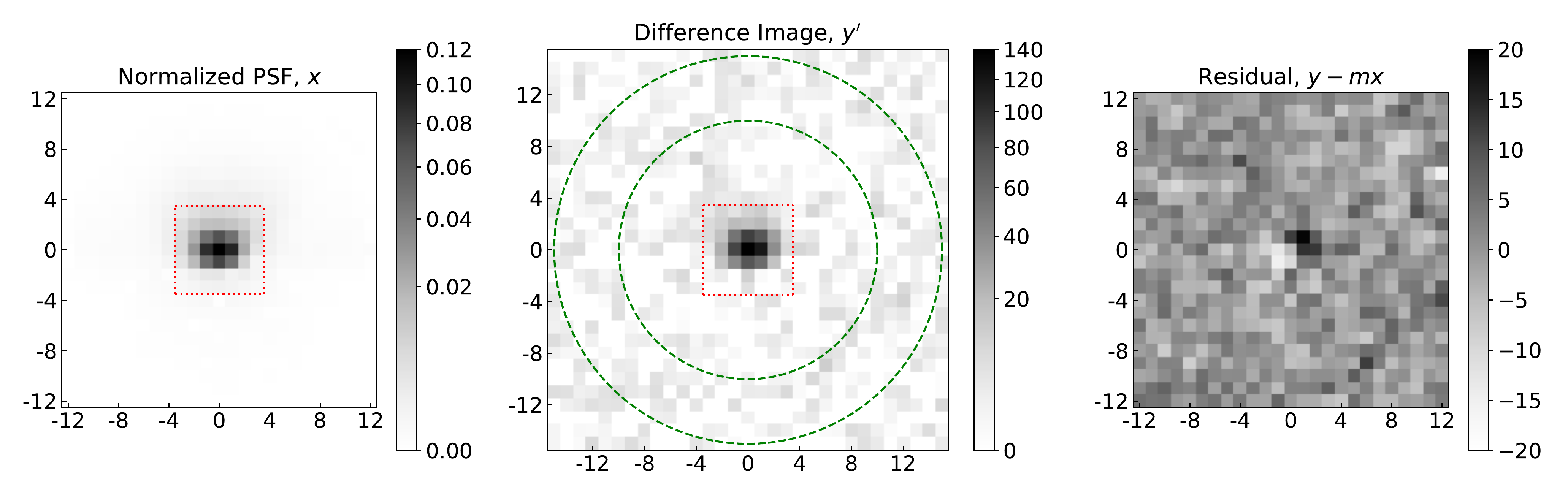}
    \caption{An example of the cutouts of the PSF-model image (left, $25\times 25$ pixels), the difference image (middle, $31\times 31$ pixels), and the residual (right, $25\times 25$ pixels) centered on the position of the target. The central $7\times 7$ pixel cutouts are marked by the dotted red squares. The background region is marked by the dashed green annulus, with inner radius = 10 pixels and outer radius = 15 pixels. Note that 10 pixels $\approx5$ FWHM.  \label{fig:2cutouts}}
\end{figure*}

\subsection{Data Description}
The pixel scale of the ZTF camera is 1.012\,\arcsec\,pixel$^{-1}$. The typical seeing-limited full-width half maximum (FWHM) of the PSF is $\sim$2\,\arcsec\,pixel$^{-1}$. Figure \ref{fig:2cutouts} shows an example of the point source cutouts of a difference image and its normalized PSF-model template. The PSF is normalized such that the sum of all pixel response values is equal to 1. These images are available at the NASA/IPAC Infrared Science Archive (IRSA)\footnote{\url{https://irsa.ipac.caltech.edu}}. 

The difference image PSF is a product of the ZOGY image subtraction algorithm \citep{Zackay2016}. ZOGY generates this by combining the input PSF templates from the science and reference images prior to subtraction. The science and reference image PSF templates were generated using an automated version of the classic DAOPhot / AllStar software \citep{Stetson1987}, with further optimizations for ZTF (\citealt{Masci2019}; Sections 3.5 \& 4). A linearly spatially-varying PSF model consisting of a Gaussian core modulated by corrections is fit to a set of pre-filtered (uncontaminated and unsaturated) stars in the science and reference images separately. Only PSF estimates at the center of the science and reference CCD-quadrants are used for input to ZOGY. Prior to use in ZOGY, the PSF templates are further regularized to suppress pixel outliers in their outer regions. ZOGY then combines the PSFs using a Fourier inversion method to generate a single PSF template for the difference image. This single PSF therefore represents an effective PSF for the entire quadrant image. It’s spatial variation on quadrant scales (inherent in the science and reference images) is $<$ 1\%. This is not significant to impact the accuracy of our PSF-fit photometry in our magnitude range of interest ($\gtrsim17$ mag), where measurements are dominated by sky background noise.

Hereafter we denote the pixel values of model image and difference image by $x_{i}$ and $y'_{i}$, respectively. $x$ is unitless and $y'$ has the unit of detector data number (DN), which is analogous to Analog Digital Units (ADU). We estimate the background noise $\sigma_{\rm bkg}$ (in the unit of DN) from all pixels inside an annulus centered at the location of the target, with an inner radius $r_{\rm in}=10$ pixels and an outer radius $r_{\rm out}=15$ pixels (indicated by the dashed green circles in Figure \ref{fig:2cutouts}). Thus, $\sigma_{\rm bkg} = 0.5 \times$ [(the 84th percentile of $\rm y'_{bkg}$) $-$ (the 16th percentile of $\rm y'_{bkg}$)]. Ideally the median of all pixels in this background annulus should be around zero, i.e., median($y'_{\rm bkg}) \approx 0$, assuming that image subtraction is perfect. However, it was found that the background level is sometimes far from zero in regions close to the center of galaxies. Therefore, we subtracted the local background from $y'$ to get a more robust estimate of the excess flux relative to background: $y_{i} = y'_{i} - {{\rm median}(y'_{\rm bkg})}$. The $y$ thus derived also has units of DN.

\subsection{The PSF Fitting Method} \label{subsec:fmcmc}
In ZSDS, image subtraction is performed using the ZOGY algorithm \citep{Zackay2016}. The PSF template image (left panel of Figure \ref{fig:2cutouts}) is generated such that the difference image (right panel of Figure \ref{fig:2cutouts}) can be modelled by \textit{the PSF image multiplied by a number $m$, plus some random noise $\epsilon$}, i.e., $y=m x+\epsilon$. Here, $m$ is the PSF-fit flux in the unit of DN, and $\epsilon$ is a noise term: $\epsilon \sim\mathcal{N} (0, \sigma^2)$. The statistical pixel uncertainty for $y_{i}$ is
\begin{equation}
\sigma_{i}^2 = \frac{y_{i}}{\rm gain} + \sigma_{\rm bkg}^2 \label{eq:sigma_i}
\end{equation}
where the gain is the electronic detector-gain (in the unit of electron per DN). 

Although our task is simply fitting a straight line to a set of ($x_i$, $y_i$) pairs, there is no consensus on how to derive the best measurement of $m$ [see \citealt{Hogg2010} or \citealt{Sharma2017} (Section 2) for a recipe on this problem]. The commonly adopted maximum likelihood estimate has the advantage of being fast, but is only optimal for the background-dominated-noise limit \citep{Zackay2016}. In principle we expect measurements of intra-night observations to be consistent with each other, but we found that our initially adopted maximum likelihood method didn't provide such a result. Instead, a Bayesian method was attempted whereby we implemented a Markov Chain Monte Carlo (MCMC) fit, which  was found to give the smallest variance of intra-night observations in the same band. Therefore, we adopted the MCMC approach, and utilized \texttt{emcee}, which is an affine invariant MCMC ensemble sampler that uses multiple walkers to sample the posterior probability distribution \citep{Goodman2010, Foreman-Mackey2013}.

Assuming the uncertainties in Eq. (\ref{eq:sigma_i}) are underestimated by a constant systematic factor $\sigma_0$, the probability of $y_{i}$ given ($x_i$, $\sigma_i$, $m$, $\sigma_0$) is
\begin{align}
 &   p(y_i | m, \sigma_0, x_i, \sigma_i) \notag \\
 =& \frac{1}{\sqrt{2\pi (\sigma_i^2 + \sigma_0^2) }}  {\rm exp}\left( - \frac{(y_i - m x_i)^2}{2(\sigma_i^2 + \sigma_0^2)}\right).\label{eqn:p_y}
\end{align}
From~\ref{eqn:p_y} it follows that the log-likelihood is:
\begin{equation}
    \ln \mathcal{L} = \sum_{i}^N \left[ { \ln}\left(\frac{1}{\sqrt{2\pi (\sigma_i^2 + \sigma_0^2)}}\right) - \left(\frac{y_i - m x_i}{2(\sigma_i^2+ \sigma_0^2)} \right)^2 \right].
\end{equation}
We only include the central $7\times 7$ cutout (indicated by the dotted red square in Figure \ref{fig:2cutouts}) in the fit, so $N=49$ is the number of pixels that were taken into consideration.  

The posterior probability distribution function of the model parameters ($m$, $\sigma_0$) for each observation can be obtained from the following equation according to Bayes' theorem:
\begin{align}
   & p(m, \sigma_0 | \{y_i\}_{i=1}^N, x_i, \sigma_i) \notag \\ 
= & \frac{1}{Z} p(\{y_i\}^N_{i=1} | m, \sigma_0, x_i, \sigma_i) p(m, \sigma_0)
\end{align}
where $Z$ is a normalization factor, and $p(m, \sigma_0)$ is the prior. 

\begin{figure}
    \centering
    \includegraphics[width=\columnwidth]{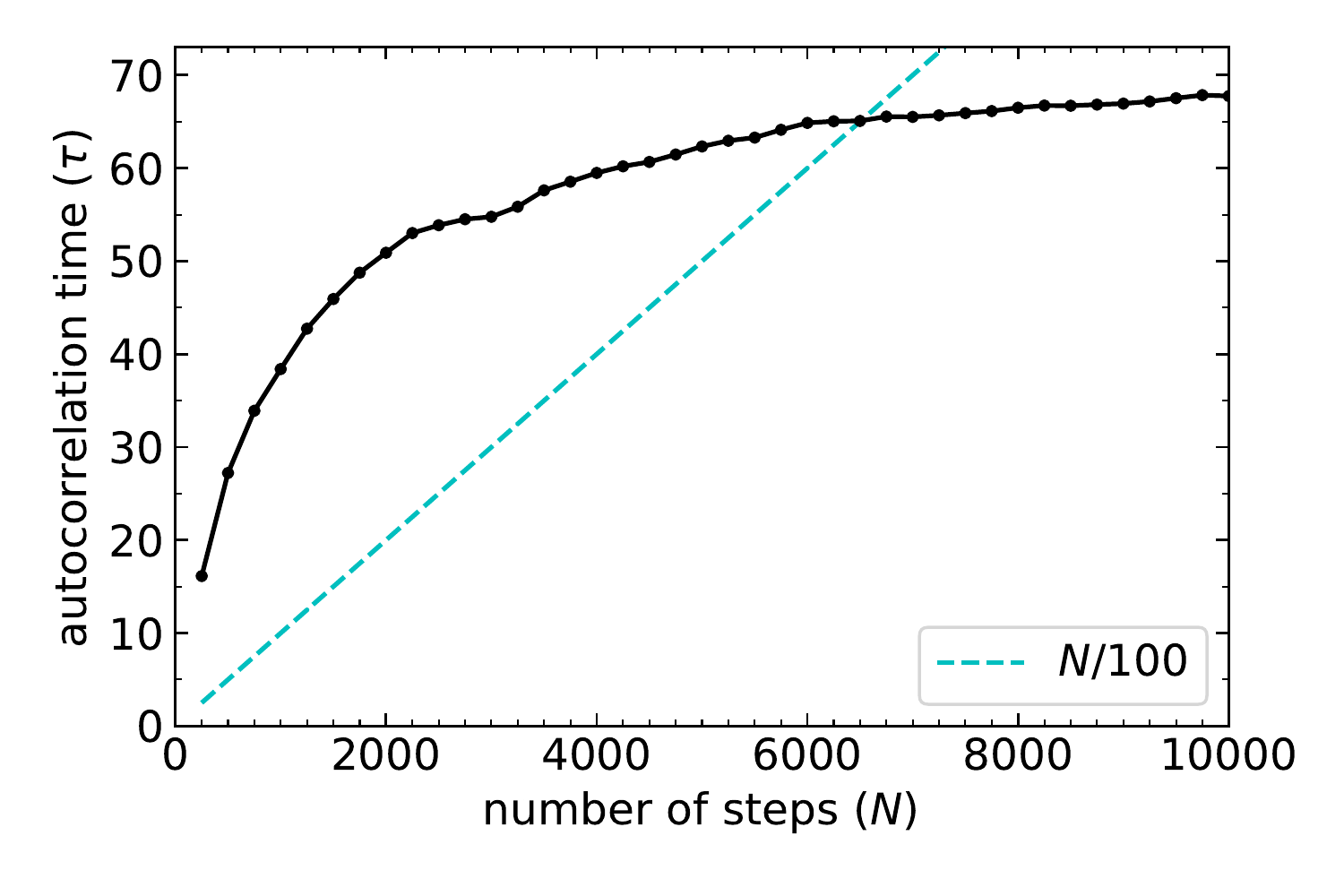}
    \caption{The solid black line shows autocorrelation time ($\tau$) as a function of sample step ($N$). We compute $\tau$ every 250 steps, and the MCMC chaines are stopped when the difference of two consecutive $\tau$ is less than 0.01. In the case shown in this figure, the MCMC chains are converged after 10,000 steps. The dashed cyan line intercepts with the black line at $N\approx 6500$. When $N>6500$, the effective number of samples is larger than 100.  \label{fig:convergence}}
\end{figure}

We adopted wide and flat priors: (i) $m$ was uniformly distributed in the range [$-10^{6}$, $10^{6}$]; (ii) $\sigma_0$ was logarithmically uniformly distributed in the range [$e^{-10}$, $e^{10}$]. The two-dimensional parameter space was investigated using 250 walkers. All models were run to convergence as determined by the evolution of the auto-correlation of the individual MCMC chains (see \url{https://emcee.readthedocs.io/en/latest/tutorials/autocorr/}). A demonstration of this step is given in Figure \ref{fig:convergence}.

We obtained the posterior probability distributions for $m$ and $\sigma_0$ as the output from the MCMC fitting, and marginalized over $\sigma_0$ to estimate the slope, $m$. Throughout this paper, we take the median value of the distribution as the measured flux, $f_{\rm mcmc}$, whereas the uncertainty on this value, $\sigma_{f_{\rm mcmc}}$, was estimated by half of the difference between the 84th and 16th percentiles of the marginalized posterior of $m$. Both $f_{\rm mcmc}$ and $\sigma_{f_{\rm mcmc}}$ have units of DN.

\begin{figure*}
    \centering
    \includegraphics[width=\textwidth]{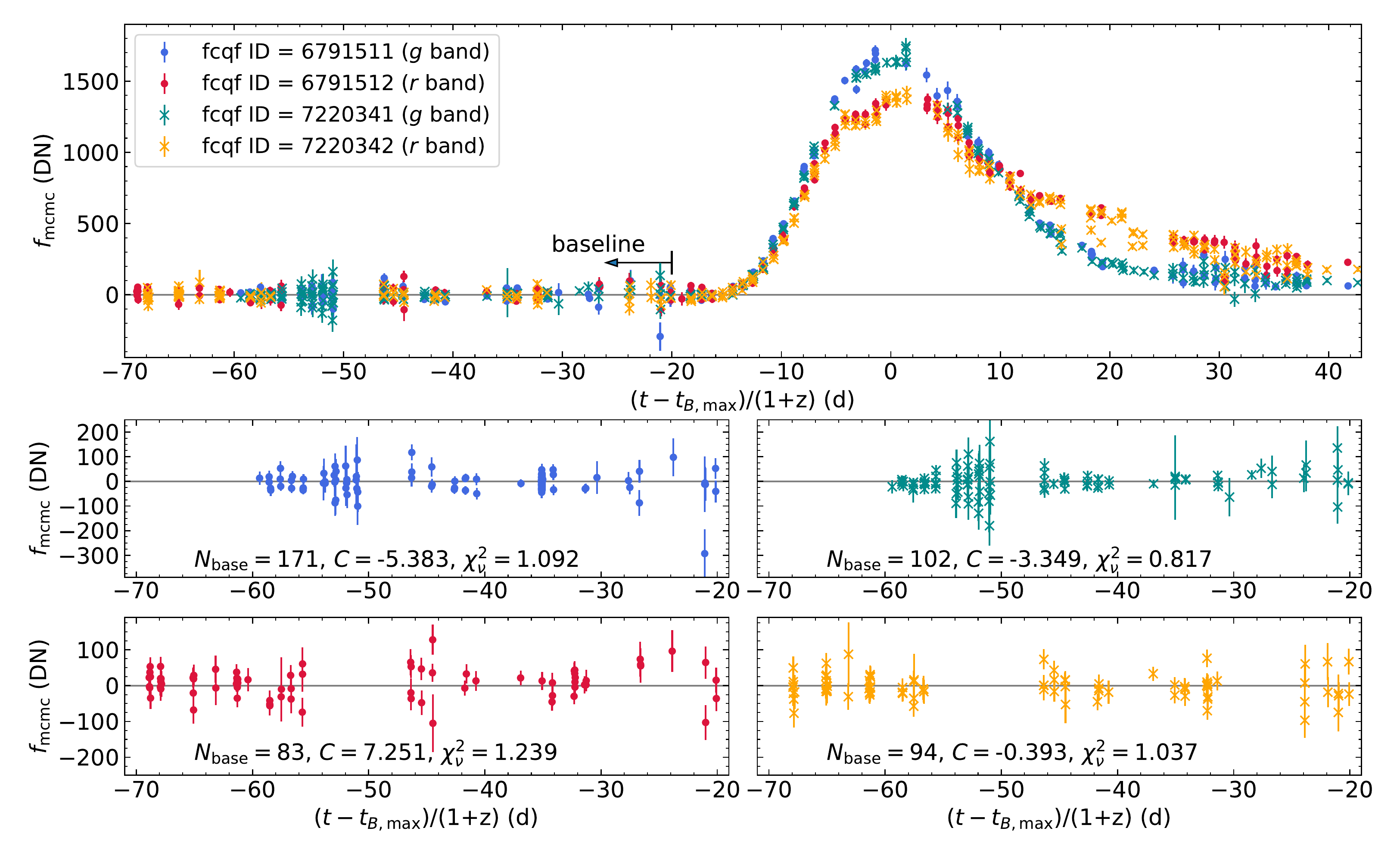}
    \caption{Upper panel: P48 light curve of ZTF18aazblzy. This target was observed in two filters ($g$ and $r$) and two fields (field 679 and 722). Observations associated with different fcqf ID (Eq. \ref{eq:fcqf}) are shown in distinctive colors. The $x$-axis shows time measured in rest-frame days relative to the  \texttt{SALT2}-estimated $B$-band maximum epoch ($t_{B, \rm max}$). Lower four panels: a zoom-in of the baseline region. The number of baseline observations ($N_{\rm base}$), the calculated offset level ($C$) and reduced chi square ($\chi^2_{\nu}$) for each fcqf ID are shown in each panel.}
    \label{fig:baseline}
\end{figure*}

\subsection{Quality Filtering}
A small fraction of the	ZTF	data were acquired through intermittent cloud	cover or featured extremely high backgrounds due to the proximity of the full moon. This affects the resulting photometric calibration. Therefore, several cuts were applied to ensure the quality of our photometric measurements.

\begin{itemize}
\item We removed data points with non-zero values of \texttt{infobits}. This keyword is a 16-bit integer that encodes the status of processing and instrumental calibration steps for the science image; specific operations that fail to meet predefined quality criteria are assigned to individual bits [0..15]. These bits can be ``AND'ed'' with a template bit-string to reject science images that failed specific calibration steps. 
\item We removed data points with \texttt{scisigpix} $> 25$. \texttt{scisigpix} is a robust estimate of spatial noise-sigma per pixel in the input science image; This is based on half the ``84.13$-$15.86'' percentile difference in pixel values. 
\item We removed data points with \texttt{seeing} $>4''$, where \texttt{seeing} captures FWHM of the point source.
\item We removed the observation if there was any pixel in the central $7\times7$ cutout with $y_i < -500$. Typically, pixels with negative values of several hundreds are caused by saturation in the reference image, and thus should be removed in the fitting. As it is difficult to define a threshold to mask out those bad pixels, we choose to remove the observations instead. 
\end{itemize}

\texttt{infobits} and \texttt{seeing} are in the header of every science image product in the archive as well as in IRSA's science image metadata database (DB) table. However, \texttt{scisigpix} is internal and not propagated to any publicly visible product, however, it can be estimated directly from the science images.

Note that while alternate prescriptions to flag observations may be adopted, we found the above cuts to be adequate to remove most of the non-photometric data in our sample. The flagged ``bad'' observations are not included in the table of final light curves accompanying this paper (Table~\ref{tab:lc}).

\subsection{Baseline Correction} \label{subsec:baseline}
A baseline correction was applied to $f_{\rm mcmc}$ to correct for any residual offset in the ``history'' of the light curve. We chose to define any data earlier than $T_{\rm before}$ days prior to $t_{B, \rm max}$ (in rest-frame) to be the ``history'' where $T_{\rm before}=20$. We visually inspected the light curves to make sure that no supernova flux was included in the baseline. For six targets (ZTF18aaykjei, ZTF18abhpgje, ZTF18abdpvnd, ZTF18aaytovs, ZTF18abddmrf, and ZTF18aawpcel) where the rest-frame rise time is obviously longer than 20 d, we adjusted the value of $T_{\rm before}$ to 25. Note that these objects are peculiar SNe with longer rise time than normal SNe Ia (See Table \ref{tab:info} and Section \ref{sec:peculiar}). Since the reference images for different fcqf\,ID (Eq. \ref{eq:fcqf}) were created by different observations, the baseline level should be determined separately for every possible combination of field and filter. For example, Figure \ref{fig:baseline} shows the light curve of ZTF18aazblzy, a normal SN Ia in our sample at redshift $z=0.0653$. The lower panels show the number of observations in each fcqf\,ID in the baseline region ($N_{\rm base}$), the offset level $C$, as well as the reduced chi square statistic ($\chi^2_{\nu}$):
\begin{equation}
    \chi^2_{\nu} = \frac{1}{\nu}\sum_{i=1}^{N_{\rm base}} \frac{ ( C- f_{\rm mcmc, i})^2 }{\sigma_{f_{\rm mcmc}, i}^2} \label{eq:chi2_red}
\end{equation}
where $\nu = N_{\rm base}-1$ is the degree of freedom and $C$ is calculated as the weighted mean of all $f_{\rm mcmc}$ measurements in the baseline. 

\begin{figure}
    \centering
    \includegraphics[width = \columnwidth]{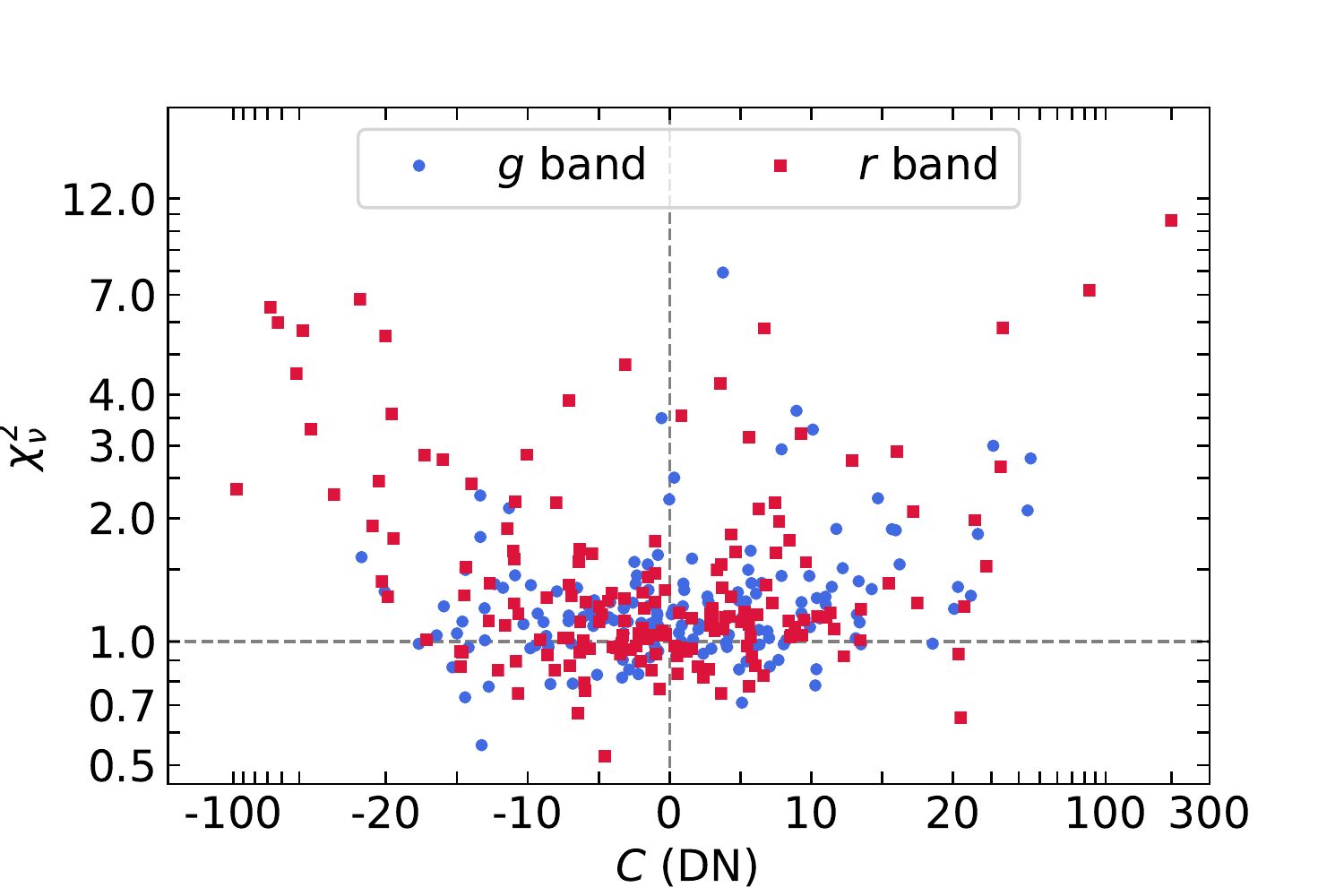}
    \caption{Distribution of $\chi^2_{\nu}$ and $C$ color coded by filters ($g$ and $r$). Note that the horizontal axis is shown with a linear scale for $-20<C<20$ and a log scale for $|C|>20$. The vertical axis is shown in a log scale. The dashed horizontal line indicates $\chi^2_{\nu}=1$ and the dashed vertical line indicates $C=0$. The median of $C$ is $-1.49$, and the median of $\chi^2_{\nu}$ is $1.21$. }
    \label{fig:offset_chi2}
\end{figure}
Figure \ref{fig:offset_chi2} shows the distribution of $\chi^2_{\nu}$ vs.~$C$ for all targets in our sample. Although we may expect $C\approx0$, a non-zero historical baseline level can occur if 
\begin{enumerate}[label=(\roman*)]
\item The reference image is contaminated by residual flux from the actual transient being measured, i.e., the input images used to construct the reference inadvertently included epochs containing significant transient flux. (This is unlikely as we applied selection step 4 in Table \ref{tab:selection}.)
\item The reference image is contaminated by an instrumental artifact that was not properly masked (or detected as an outlier) prior to co-addition.
\item There are systematic residuals from \textit{persistently} inaccurate gain-matching between the science and reference images. This is usually triggered by imperfect flat-fielding of the science images used to construct the reference image, i.e., the reference image exhibits a spatial variation in its photometric gain. This systematic spatial variation will persist (be imprinted) in \textit{all} subtraction images constructed using this reference image. If the gain-mismatch between science and reference images at the location of the transient is significant, this will also lead to inflated $\chi^2_{\nu}$ values since measurements \textit{relative to the baseline} will be noisier (inflated by a hidden systematic gain factor) than those represented by the individual-epoch measurement uncertainties ($\sigma_{f_{\rm mcmc}, i}$).
\end{enumerate}

A sufficient number of historical measurements is required for robust estimates of $C$ and $\chi^2_{\nu}$. Large absolute values of $C$ or large values of $\chi^2_{\nu}$ should be considered as ``red flags'' suggesting further analysis and visual examination of the images, particularly the reference image to search for the systematic described in case (iii) above. We mitigate this systematic by subtracting the baseline $C$ from the measured $f_{\rm mcmc}$, and if $\chi^2_{\nu}>1$, we multiplied the raw $\sigma_{f_{\rm mcmc}}$ by $\sqrt{\chi^2_{\nu}}$. The photometric uncertainties thus derived should be considered as a conservative estimate. For others who would like to model these light curves in the future, it is also advised to perform such a baseline validation and uncertainty scaling, or to remove observations associated with $\chi^2_{\nu}\gtrsim4$ or $|C|\gtrsim15$ from the sample. In the light curves accompanying this paper (Table \ref{tab:lc}) we provide our measurements of $C$ and $\sqrt{\chi^2_{\nu}}$, and set the values of these columns to $-999$ if the corresponding fcqf\,ID has $N_{\rm base}< 30$. 

With the knowledge of difference image zero point magnitude (zp) provided by ZSDS, the zero-point flux in the unit of DN ($f_0$) can be calculated:
\begin{subequations}
\begin{align}
    f_0 =& 10^{0.4\times {\rm zp}}\\
    \sigma_{f_0} =&f_0 \frac{\rm ln(10)}{2.5} \sigma_{\rm zp}
\end{align}
\end{subequations}
Thus, the dimensionless flux ratio is:
\begin{subequations}
\begin{align}
    f_{\rm ratio} =& f_{\rm mcmc} / f_0\\
    \sigma_{f_{\rm ratio}} =& \sqrt{ \left(\frac{\sigma_{f_{\rm mcmc}}}{f_0}\right)^2 + \left(\frac{f_{\rm mcmc}\sigma_{f_0}}{f_0^2}\right)^2  }
\end{align}
\end{subequations}

We define the detection threshold of signal-to-noise ratio (SNT) to be 3.\footnote{We follow the rationale illustrated in \url{http://web.ipac.caltech.edu/staff/fmasci/ztf/forcedphot.pdf} to choose $\mathrm{SNT}=3$ and use $5\times \sigma_{f_{\rm ratio}}$ to define upper limits.} That is to say, whenever $f_{\rm ratio}>3\times \sigma_{f_{\rm ratio}}$, the conversion from flux to magnitude can be applied:
\begin{subequations}
\begin{align}
    m =& -2.5 \times \log f_{\rm ratio}\\
    \sigma_{m-} = & ~2.5\times \log(1 + \sigma_{f_{\rm ratio}}/f_{\rm ratio}) {\; \rm brighter\,end}\\
    \sigma_{m+} = & -2.5\times{\log}(1 - \sigma_{f_{\rm ratio}}/f_{\rm ratio}){\; \rm fainter\,end}
\end{align}\label{eq:sigma_m}
\end{subequations}
For non-detections, we compute 5$\sigma$ upper limits as
\begin{equation}
    m_{\rm lim} = -2.5 \times {\log}(5\times \sigma_{f_{\rm ratio}}). \label{eq:upplim}
\end{equation}

\subsection{Final Sample Selection} \label{subsec:final sample selection}
With the 154 targets selected in Section \ref{subsec:initial sample selection}, we further apply the cuts illustrated in steps 5--7 of Table~\ref{tab:selection}. Steps 5 and 6 are made to ensure that targets included in our sample have both early time detections and relatively dense light curve sampling. The requirement in step 7 is to ensure that epochs of maximum light in $g$ and $r$ can be accurately estimated (in Section \ref{subsec:lc_fitting}). In the end, 127 SNe Ia were finally retained in our sample. Table \ref{tab:info} provides general information on these targets. Table \ref{tab:phot} summarizes additional photometric properties of this sample, and Table \ref{tab:lc} provides their forced-PSF photometry light curves. Photometric and spectroscopic observations of sources rejected by our sample selection criteria will be published in a separate study on the full sample of SNe Ia found by ZTF.

\begin{deluxetable*}{lccclccl}
\tabletypesize{\scriptsize}
\tablecaption{General information of 127 SNe Ia.\label{tab:info}}
\tablehead{
\colhead{ZTF Name}   
& \colhead{R.A. (J2000)} 
& \colhead{Decl. (J2000)} 
& \colhead{IAU} 
& \colhead{TNS Intermal}  
& \colhead{Telescope} 
& \colhead{Spec Phase} 
& \colhead{Ia Subtype}  \\
\colhead{(ZTF18)}              
& \colhead{(degree)}     
& \colhead{(degree)}      
& \colhead{Name}     
& \colhead{Name}   
& \colhead{}  
& \colhead{}   
& \colhead{}     
}
\decimalcolnumbers
\startdata
aansqun  & 251.2920505  &  42.7178671   & SN2018dyp & ZTF18aansqun & P60        & $-$1  & normal \\
aapqwyv  & 251.6054101  &  25.4214872   & SN2018bhc & Gaia18bek    & DCT        & $+$0  & normal$\ast$ \\
aaqcozd  & 190.5630550  &  42.2717146   & SN2018bjc & ATLAS18odd   & NOT        & $-$4  & normal \\
aaqqoqs  & 207.9703694  &  47.2569927   & SN2018cbh & ZTF18aaqqoqs & P200       & $-$6  & 99aa-like \\
aawpcel  & 195.3210357  &  59.8100342   & SN2018cir & ZTF18aawpcel & P60        & $+$0  & SC$\ast$ \\
aaxakhh  & 278.5608703  &  32.9193280   & SN2018cvb & PS18aca      & P60        & $+$29         & 91T-like$\ast$ \\
aaykjei  & 244.9123513  &  49.1851778   & SN2018crl & ZTF18aaykjei & P60,P60,APO,P60,P200       & $+$3,$-$4,$-$10,$+$14,$+$81   & Ia-CSM \\
aaytovs  & 266.4724817  &  31.7105165   & SN2018crk & ZTF18aaytovs & P200       & $+$14         & 99aa-like \\
aazblzy  & 242.8394794  &  36.9943003   & SN2018cri & ZTF18aazblzy & P200       & $-$10         & normal \\
abauprj  & 254.7709972  &  47.2364390   & SN2018cnw & ZTF18abauprj & NOT        & $-$6  & 99aa-like \\
abclfee  & 258.5925379  &  48.2643039   & SN2018cxk & ZTF18abclfee & P60,LT,P200,P60,P200       & $+$2,$+$4,$+$8,$+$8,$+$35     & 02cx-like \\
abddmrf  & 226.9673020  &  38.0484486   & SN2018dsx & ATLAS18sdi   & P200       & $+$37         & SC$\ast$ \\
abdmgab  & 250.9022676  &  33.5336160   & SN2018lph & ZTF18abdmgab & Keck1      & $+$1  & 86G-like \\
abdpvnd  & 348.5225939  &  29.6333864   & SN2018dvf & ZTF18abdpvnd & P60,P60,NOT        & $-$7,$+$31,$+$59      & SC \\
abgmcmv  & 274.0545029  &  55.5908773   & SN2018eay & ZTF18abgmcmv & Keck1      & $-$16         & 91T-like \\
abhpgje  & 277.6781453  &  54.6344776   & SN2018eul & ATLAS18tje   & P60,P200   & $+$12,$+$67   & SC \\
abptsco  & 261.7142166  &  34.2465793   & AT2018lpm & ZTF18abptsco & from Atel 12052    & ?     & normal$\ast$ \\
abuqugw  & 244.5576659  &  39.1238139   & SN2018geo & ATLAS18vca   & P60        & $-$9  & normal \\
\enddata
\tablecomments{
Column (4): SN IAU name from TNS.
Column (5): TNS internal name (indicating the discovery group).
Column (6) \& (7): Follow-up telescope and epoch of the spectrum with respect to $t_{g, \rm max}$ used to determine the spectral subtype. $t_{g, \rm max}$ is the epoch of $g$-band maximum light provided in Table \ref{tab:phot} and calculated in Section \ref{subsec:lc_fitting}.
Column (8): If classification can not be \textit{reliably} determined from spectroscopy alone, the subtype ends with an asterisk. See Section \ref{subsec:subtype} for details.
This table is available in its entirety in the machine-readable form.}
\end{deluxetable*}
\begin{deluxetable*}{lllcccc}[ht!]
\tabletypesize{\scriptsize}
\tablecaption{Photometric properties of 127 SNe Ia.\label{tab:phot}}
\tablehead{
\colhead{ZTF Name}   
& \colhead{Redshift} 
& \colhead{$E(B-V)$}
& \colhead{$t_{g, \rm max}$}  
& \colhead{$\Delta m_{15}(g)$} 
& \colhead{\texttt{SALT2} $x_1$}
& \colhead{\texttt{SALT2} $c$}\\
\colhead{(ZTF18)}   
& \colhead{}   
& \colhead{(mag)}
& \colhead{(MJD)}   
& \colhead{(mag)}   
& \colhead{}   
& \colhead{}  \\
\colhead{(1)}  
& \colhead{(2)}  
& \colhead{(3)}  
& \colhead{(4)}  
& \colhead{(5)}  
& \colhead{(6)}  
& \colhead{(7)} 
}
\startdata
aansqun & 0.0597        & 0.0127    & 58313.56 $\pm$ 0.18 & 0.99 $\pm$ 0.08 & -1.72 $\pm$ 0.61 & 0.20 $\pm$ 0.10 \\
aapqwyv & 0.0560        & 0.0390    & 58243.79 $\pm$ 0.18 & 0.99 $\pm$ 0.04 & -1.72 $\pm$ 0.18 & 0.17 $\pm$ 0.04 \\
aaqcozd & 0.0732        & 0.0203    & 58253.02 $\pm$ 0.06 & 0.96 $\pm$ 0.02 & -1.23 $\pm$ 0.07 & -0.10 $\pm$ 0.01 \\
aaqqoqs & 0.080         & 0.0137    & 58261.24 $\pm$ 0.03 & 0.68 $\pm$ 0.03 & 1.23 $\pm$ 0.27 & -0.01 $\pm$ 0.03 \\
aawpcel & 0.150         & 0.0108    & 58279.48 $\pm$ 1.43 & 0.53 $\pm$ 0.09 & 3.24 $\pm$ 0.78 & 0.12 $\pm$ 0.06 \\
aaxakhh & 0.120         & 0.0586    & 58266.65 $\pm$ 0.09 & 0.75 $\pm$ 0.03 & 0.54 $\pm$ 0.22 & -0.08 $\pm$ 0.02 \\
aaykjei & 0.0970        & 0.0137    & 58291.72 $\pm$ 0.35 & 0.44 $\pm$ 0.03 & 4.14 $\pm$ 0.21 & 0.50 $\pm$ 0.02 \\
aaytovs & 0.0746        & 0.0523    & 58293.58 $\pm$ 0.03 & 0.59 $\pm$ 0.03 & 2.03 $\pm$ 0.37 & 0.12 $\pm$ 0.04 \\
aazblzy & 0.0653        & 0.0116    & 58291.47 $\pm$ 0.02 & 1.02 $\pm$ 0.02 & -1.68 $\pm$ 0.09 & -0.07 $\pm$ 0.02 \\
abauprj & 0.0242        & 0.0230    & 58301.59 $\pm$ 0.02 & 0.67 $\pm$ 0.01 & 1.34 $\pm$ 0.04 & -0.01 $\pm$ 0.01 \\
abclfee & 0.0290        & 0.0127    & 58299.13 $\pm$ 0.16 & 1.67 $\pm$ 0.04 & -2.53 $\pm$ 0.09 & 0.19 $\pm$ 0.02 \\
abddmrf & 0.164         & 0.0119    & 58307.57 $\pm$ 1.02 & 0.48 $\pm$ 0.10 & 2.53 $\pm$ 0.94 & 0.08 $\pm$ 0.05 \\
abdmgab & 0.0803        & 0.0186    & 58310.98 $\pm$ 0.18 & 1.08 $\pm$ 0.04 & -2.31 $\pm$ 0.33 & 0.12 $\pm$ 0.05 \\
abdpvnd & 0.050         & 0.0907    & 58319.07 $\pm$ 0.28 & 0.60 $\pm$ 0.03 & 3.06 $\pm$ 0.10 & 0.15 $\pm$ 0.01 \\
abgmcmv & 0.0185        & 0.0397    & 58328.52 $\pm$ 0.01 & 0.68 $\pm$ 0.01 & 0.69 $\pm$ 0.05 & 0.63 $\pm$ 0.01 \\
abhpgje & 0.1342        & 0.0319    & 58333.86 $\pm$ 0.74 & 0.50 $\pm$ 0.04 & 3.48 $\pm$ 0.34 & -0.01 $\pm$ 0.03 \\
abptsco & 0.12  & 0.0269    & 58363.13 $\pm$ 0.37 & 0.84 $\pm$ 0.10 & -0.40 $\pm$ 0.83 & 0.07 $\pm$ 0.07 \\
abuqugw & 0.0313        & 0.0064    & 58384.85 $\pm$ 0.04 & ... & -1.35 $\pm$ 0.03 & -0.12 $\pm$ 0.01 \\
\enddata
\tablecomments{
Column (2): The host galaxy redshift ($z$) is shown with 4 decimals if: (i) $z$ is taken from NED, (ii) $z$ is measured from the galaxy spectrum obtained by us, (iii) $z$ is measured from a SN spectrrum where the host H$\alpha$ line can be identified. Redshifts inferred from \texttt{SNID} fit on SN spectra are shown with 3 decimals. Redshift of ZTF18abptsco was reported by ATel 12052 \citep{Gomez2018} and was shown with 2 decimals. 
Column (3): Forground Galactic extinction from \citet{Schlafly2011}.
Column (4) \& (5): Values of $t_{g, \rm max}$ and $\Delta m_{15}(g)$ for peculiar events (classified as ``02cx-like'', ``Ia-CSM'', ``SC'' or ``SC$\ast$'') were obtained from polynomial fits to the $g$-band light curve. For other objects, both $t_{g, \rm max}$ and $\Delta m_{15}(g)$ were from \texttt{SALT2} estimates. 
If a target does not have any $g$-band observations in the time range [$t_{g, \rm max}$+10, $t_{g, \rm max}$+20], $\Delta m_{15}(g)$ can not be securely estimated and is thus shown in blank (e.g., ZTF18abuqugw).
Column (6) \& (7): the light curve shape parameter $x_1$ and color $c$ from \texttt{SALT2} fitting.
This table is available in its entirety in the machine-readable form.
}
\end{deluxetable*}
\begin{deluxetable*}{lrlrrrl}
\tabletypesize{\normalsize}
\tablecaption{Samples of low-to-intermediate redshift SNe Ia\label{tab:surveys}}
\tablehead{
\colhead{Catalog} 
& \colhead{Size} 
& \colhead{Sample Redshift} 
& \colhead{$N_\mathrm{early}$\tablenotemark{a}} 
& \colhead{Obs nights} 
& \colhead{Time span} 
& \colhead{Bands}
}
\decimalcolnumbers
\startdata
CfA3 & 185  & 0.01--0.085 (0.027) & 9  & 12 &  2001--2008&  $BVRIr'i'$\\
CfA4 & 94 &  0.0055--0.073 (0.029) & 3 & 16 &2006--2011 &($u'U$)$BVr'i'$ \\
CSP-I & 134 & 0.0037--0.0835 (0.0241) & 14 & 28 & 2004--2009 &  $uBgVriYJH$\\
CSP-II & 214 & 0.0036--0.1376 (0.0341) & --  & -- & 2011--2015 &$uBgVriYJH$ \\
Foundation-I & 225 & 0.004--0.11 (0.033) &23  &  7 & 2015--2017 & $grizy_{\rm P1}$\\
\hline
LOSS & 165 & 0.002--0.095 (0.0194) & 32  & 21  & 1998--2008 & $BVRI$\\
SDSS-II\tablenotemark{b} & 327 & 0.037--0.4 ($\sim0.21$) &72 & 9 & 2005-2007 & $ugriz$  \\
PTF/iPTF & 265 & 0.00068--0.19 (0.083) & \textbf{138}  &20 & 2009--2014  & $R$ \\
TESS-2018  & 18  & 0.0163--0.09 (0.04) & 16  & 18  & 2018 &$i_{\rm TESS}$ \\
ZTF-2018 & \textbf{336} & 0.01815--0.164 (0.074) & 127  & \textbf{43} & 2018 & $gr$ \\
\enddata
\tablecomments{
Column (2): Total number of objects in each catalog.
Column (3): Redshift range with the median value shown in parenthesis. 
Column (4): Number of targets with early observations.
Column (5): Median number of nights of observation per SN (upper limits not included). 
Column (6): Survey period.
Column (7): Observing bands, those shown in parenthesis means less than half of targets were observed with the corresponding bands. \\
Some estimates can not be made due to our limited access to data. 
\\
References: the Harvard-Smithsonian Center for Astrophysics SNe Ia sample (CfA3, \citealt{Hicken2009}; CfA4 \citealt{Hicken2012}), the Carnegie Supernova Project I (CSP-I) low-redshift sample \citep{Hamuy2006, Contreras2010, Stritzinger2011, Krisciunas2017}, CSP-II \citep{Phillips2019}, the Foundation Supernovae Survey data release I \citep{Foley2018}, the Sloan Digital Sky Survey II Supernova Survey (SDSS-II; \citealt{Frieman2008, Sako2018}), the Lick Observatory Supernova Search (LOSS, \citealt{Filippenko2001}) follow-up photometry program \citep{Ganeshalingam2010}, the (intermediate) Palomar Transient Factory (PTF/iPTF, \citealt{Papadogiannakis2019}), and SNe Ia observed in the first six sectors of the Transiting Exoplanet Survey Satellite (TESS) with pre-explosion observations \citep{Fausnaugh2019}. }
\tablenotetext{a}{The number of SNe Ia with early observations, $N_\mathrm{early}$, is defined as the number of targets with detections earlier than 10\,d in rest-frame relative to the epoch of $B$-band peak luminosity. For SDSS-II this estimate is made relative to the time of $g$-band maximum light, while for the CfA3 and LOSS samples we use the observer-frame phase.}
\tablenotetext{b}{The size of the SDSS-II sample was reported to be 327 in \citep{Frieman2008}, and increased to $\sim$500 later \citep{Hayden2010a, Hayden2010b}. \citet{Sako2018} claimed 1364 SNe Ia from SDSS with spectroscopic redshifts (though some of these SNe were identified photometrically). The statistics shown for the SDSS-II sample are based on \citet{Frieman2008}.}
\end{deluxetable*}

\section{Sample Properties} \label{sec:properties}
We use ZTF transient names throughout this paper. In some cases these events were first reported by other groups [see column (5) of Table~\ref{tab:info}], although the first detection in their forced photometry light curves may precede the time of announcement on the Transient Name Server (TNS).

\subsection{Comparison with Previous Samples}
\begin{figure}[htbp!]
    \centering
    \includegraphics[width=0.9\columnwidth]{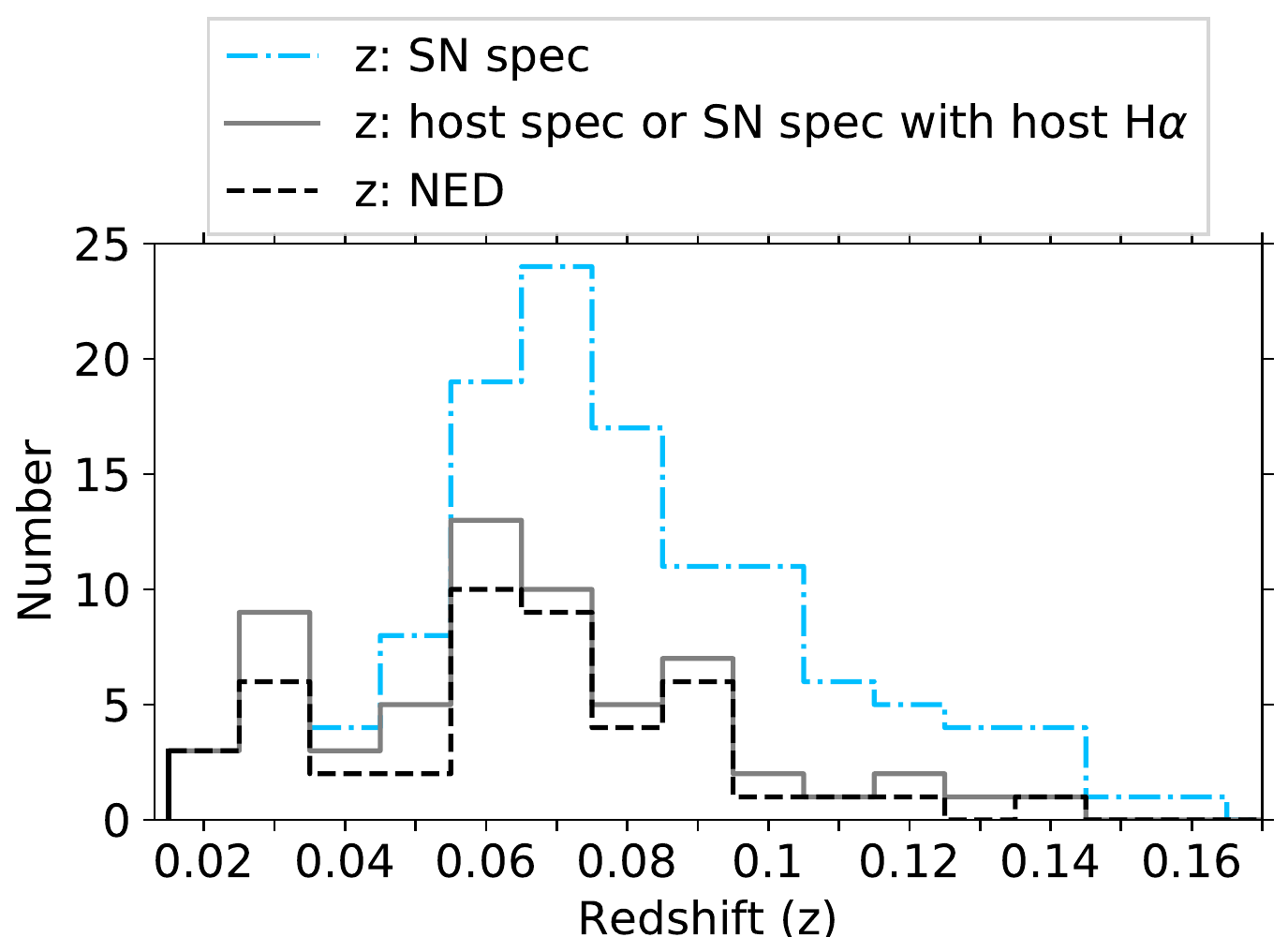}
    \caption{Redshift distribution of 127 SNe Ia in our sample. Note that this is a stacked histogram, such that the blue dot-dashed line is the total distribution. \label{fig:redshift_distribution}}
\end{figure}

The redshift distribution of all targets in our sample is plotted in Figure~\ref{fig:redshift_distribution}. Shown in black are 46 SNe with host galaxy redshifts from the NASA/IPAC Extragalactic Database (NED), while redshifts of the 16 shown in grey are measured from host galaxy lines in the SN spectrum or from a spectrum of the host galaxy itself. Redshifts of the remaining 65 targets shown in blue are inferred from the SN spectrum (see Section~\ref{subsec:subtype} for details). Redshifts estimated directly from the SN spectrum have a typical uncertainty of $\sim$0.004 (Fremling et al., 2019, in prep.). The 16th, 50th, and 84th percentiles of the redshift distribution are $z=0.051$, 0.074, and 0.106. 

We summarize early-time photometric samples of low-to-intermediate redshift SNe Ia from multiple surveys in Table~\ref{tab:surveys}. The table is divided in two, with the top half listing surveys focused on the follow-up of SNe Ia, while the bottom half lists surveys that both discover \textit{and} follow-up SNe Ia. We use this split to better highlight the number of SNe with early observations, because these numbers are not directly comparable between surveys that discover SNe and those that only perform follow-up observations. Hereafter we only compare the ZTF sample with the LOSS, SDSS-II, iPTF/PTF, and \textit{TESS} samples.

\begin{figure}[htbp!]
    \centering
    \includegraphics[width=\columnwidth]{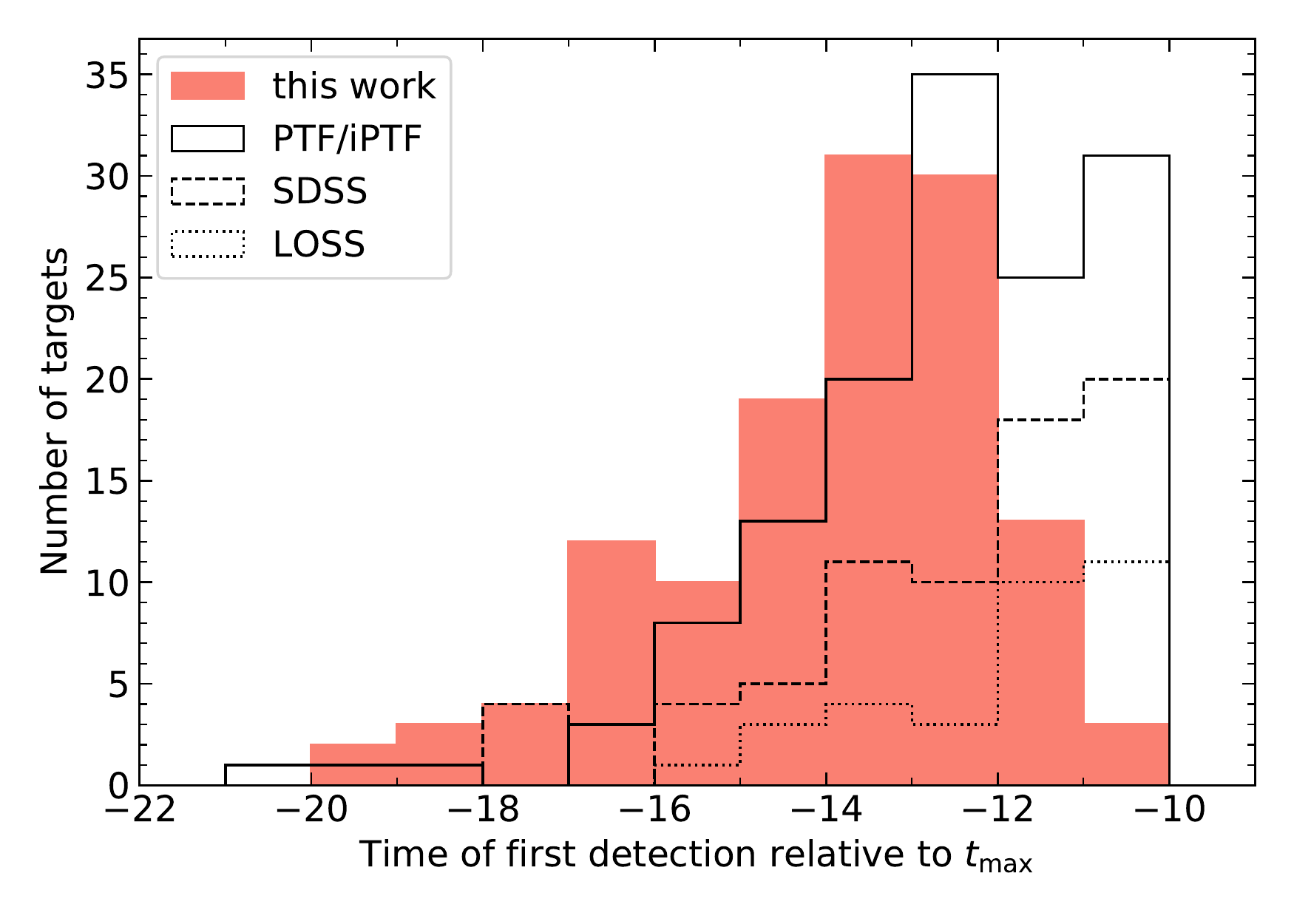}
    \caption{Histogram of first detection epoch relative to$B$ or $g$ band maximum light for ZTF, PTF/iPTF, SDSS, and LOSS. Times are given in the rest-frame except for the LOSS sample, where we use the observer frame instead. The x-axis is truncated at $-10$\,d. For the ZTF sample, the median is $-13.6$\,d and the mean is $-13.9$\,d.\label{fig:t1det2peak}}
\end{figure}

\begin{figure}[htbp!]
    \centering
    \includegraphics[width=\columnwidth]{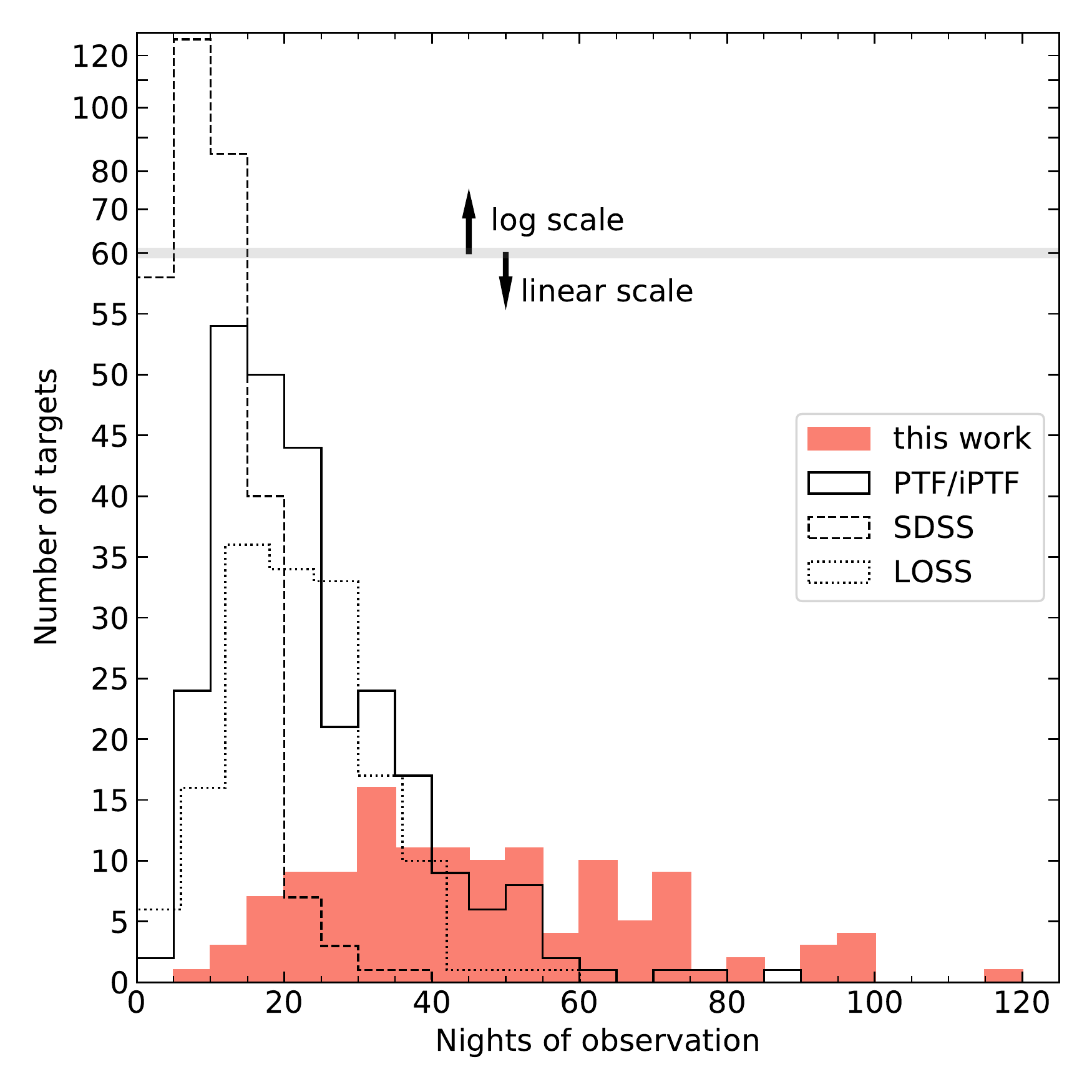}
    \caption{Histogram of the number of nights each SN Ia was observed. For the ZTF sample, the median is 43 nights and the mean is 46 nights. \label{fig:ndetections}}
\end{figure}

Figure \ref{fig:t1det2peak} shows the distribution of first detection epoch relative to $B$ or $g$ band maximum light for ZTF, PTF/iPTF, SDSS, and LOSS. SNe with no pre-maximum detections are not shown in the figure. Figure \ref{fig:ndetections} shows the histogram of the number of nights that each target was observed (upper limits are not included, intra-night observations are counted as 1 night). The \textit{TESS}-2018 sample is not plotted because it has a relatively small size. We note that column (5) of Table~\ref{tab:surveys} may not be appropriate for this sample, since \textit{TESS} is a space satellite that provides 30\,min-cadence light curves \citep{Ricker2015}. Among the 18 \textit{TESS} SNe, 7 were observed in two sectors and 11 were covered in one sector (27\,d per sector). Ten events were detected at least 14\,d prior to maximum light (see Figure 1 of \citealt{Fausnaugh2019}).

LOSS is a targeted survey that uses the Katzman Automatic Imaging Telescope (KAIT, \citealt{Filippenko2001, Li2003}) as its discovery engine. As shown in Figure \ref{fig:t1det2peak}, this sample contains the smallest number of events with $<-12$\,d observations, likely as a result of KAIT's relatively small aperture ($m_\mathrm{lim} \approx 19$\,mag) and slower cadence ($\sim$3.5\,d).

The SDSS-II survey provides one of the largest samples, which has a higher median redshift due to its large aperture ($m_\mathrm{lim} \approx 22.2$\,mag). However, Figure \ref{fig:ndetections} implies that the intervals between consecutive observations are relatively large due to the 4.5\,d cadence.

PTF/iPTF observations were carried out at a variety of cadences between 1\,d and 4\,d. Although the number of events with early-time photometry is comparable to the ZTF sample (Figure \ref{fig:t1det2peak}), only 27 (4) objects have more than 40 (60) nights of observations (Figure \ref{fig:ndetections}). Furthermore, the PTF/iPTF photometry is only in $R$ band. In comparison, from the first 7 months of the ZTF high-cadence experiment, the number of objects with 40-or-more nights of observation is 71, and the number of objects with 60-or-more nights of observation is 35. Only 11 targets were observed on less than 20 nights.

Among the 127 ZTF SNe Ia in this study, 50 were discovered at least 14\,d prior to $t_{B, \rm max}$, with 9 being detected $>17$\,d before $t_{B, \rm max}$. Among the latter 9 events, three are peculiar events with longer rise time than normal SN Ia (ZTF18aaykjei, ZTF18abdpvnd, and ZTF18abhpgje; see Section \ref{sec:peculiar} for details), four are at very low redshift (ZTF18aasdted, ZTF18abcflnz, ZTF18abfhryc, and ZTF18abauprj, all within $z=0.04$); and two have possible early-time flux excess (ZTF18abxxssh at $z=0.064$ and ZTF18aavrwhu at $z=0.062$, a detailed analysis of these two  will be presented elsewhere). In short, it is the rich information contained in the multi-band, well-sampled, early light curves that distinguishes ZTF as a unique survey for the study of early-time SNe Ia. 


\subsection{Subtype Classification}\label{subsec:subtype}
The majority of SNe Ia follow a light-curve width vs. peak luminosity relation \citep{Phillips1993}. Smaller sub-classes can be characterized by their peculiar spectroscopic and photometric properties. For example, the overluminous 91T-like events \citep{Fillippenko1992a} and 99aa-like events \citep{Li2001} have distinct \ion{Fe}{III} lines dominating their early spectra; while the subluminous 91bg-like events \citep{Filippenko1992b} and 86G-like events \citep{Phillips1987} display pronounced \ion{Si}{II} and \ion{Ti}{II} lines. Recent reviews of the observational characteristics and physical interpretation of different subtypes can be found in the literature (e.g., \citealt{Parrent2014, Maeda2016, Taubenberger2017}).

\subsubsection{Spectroscopic Observations}\label{subsubsec:specobs}
A large fraction of our spectroscopic follow-up observations were conducted by the Spectral Energy Distribution Machine (SEDM, \citealt{Blagorodnova2018}, \citealt{Rigault2019}) on the robotic Palomar 60-inch telescope (P60, \citealt{Cenko2006}). Other follow-up instruments include the Double Spectrograph (DBSP; \citealt{Oke1982}) on the Palomar 200-inch telescope (P200), the Low-Resolution Imaging Spectrometer (LRIS; \citealt{Oke1995}) on the Keck I 10-m telescope, the Andalucia Faint Object Spectrograph and Camera (ALFOSC) on the 2.56-m Nordic Optical Telescope (NOT), the Dual Imaging Spectrograph (DIS) on the Astrophysical Research Consortium (ARC) 3.5-m telescope at Apache Point Observatory (APO), the SPectrograph for the Rapid Acquisition of Transients (SPRAT) on the 2.0-m Liverpool Telescope (LT), and the Deveny spectrograph on the 4.3-m Discovery Channel Telescope (DCT; \citealt{Bida2014}). 

\begin{figure}[htbp!]
    \centering
    \includegraphics[width=\columnwidth]{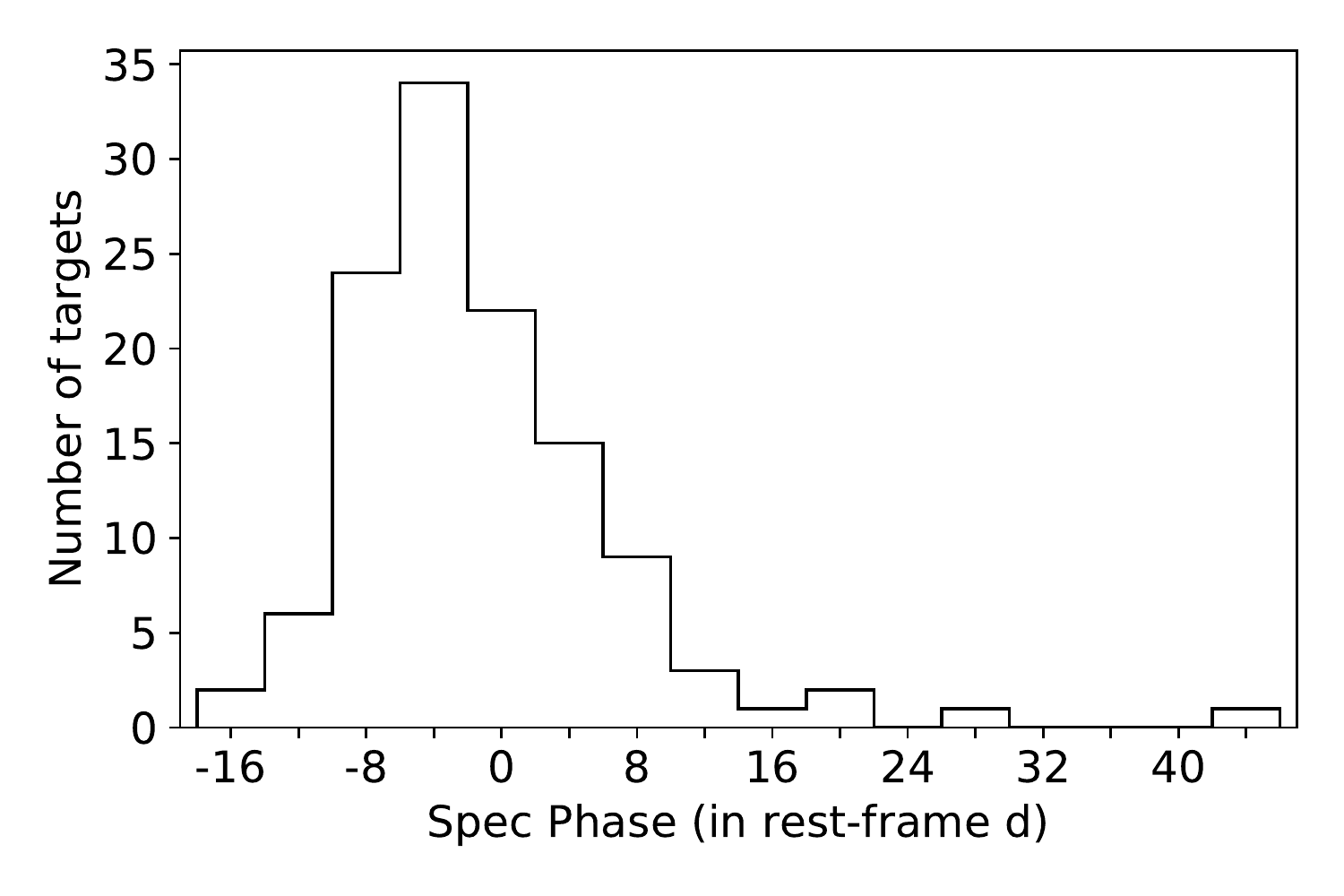}
    \caption{Histogram of the phase of the spectrum with which the classification was made for 120 targets in our sample. The median is $-3$\,d. ZTF18abptsco and the six peculiar objects are not included. \label{fig:dis_specphase}}
\end{figure}

Within our sample, 95 objects have 1 spectrum; 22 have two spectra, 5 have 3 spectra, 1 has 4 spectra, 2 have 5 spectra, and 1 has 10 spectra. The classification of ZTF18abptsco was reported via The Astronomer's Telegram \citep{Gomez2018}. If an object has more than one spectrum, we choose the one of highest signal-to-noise ratio or closest to maximum light for classification. The spectral phase for individual SNe is shown in Column (7) of Table \ref{tab:info}. Figure \ref{fig:dis_specphase} shows the distribution of rest-frame epoch relative to $g$-band maximum light (the estimation of $t_{g, \rm max}$ is illustrated in Section~\ref{subsec:lc_fitting}) used to determine the spectral subtype. 

\begin{figure}[htbp!]
    \centering
    \includegraphics[width=\columnwidth]{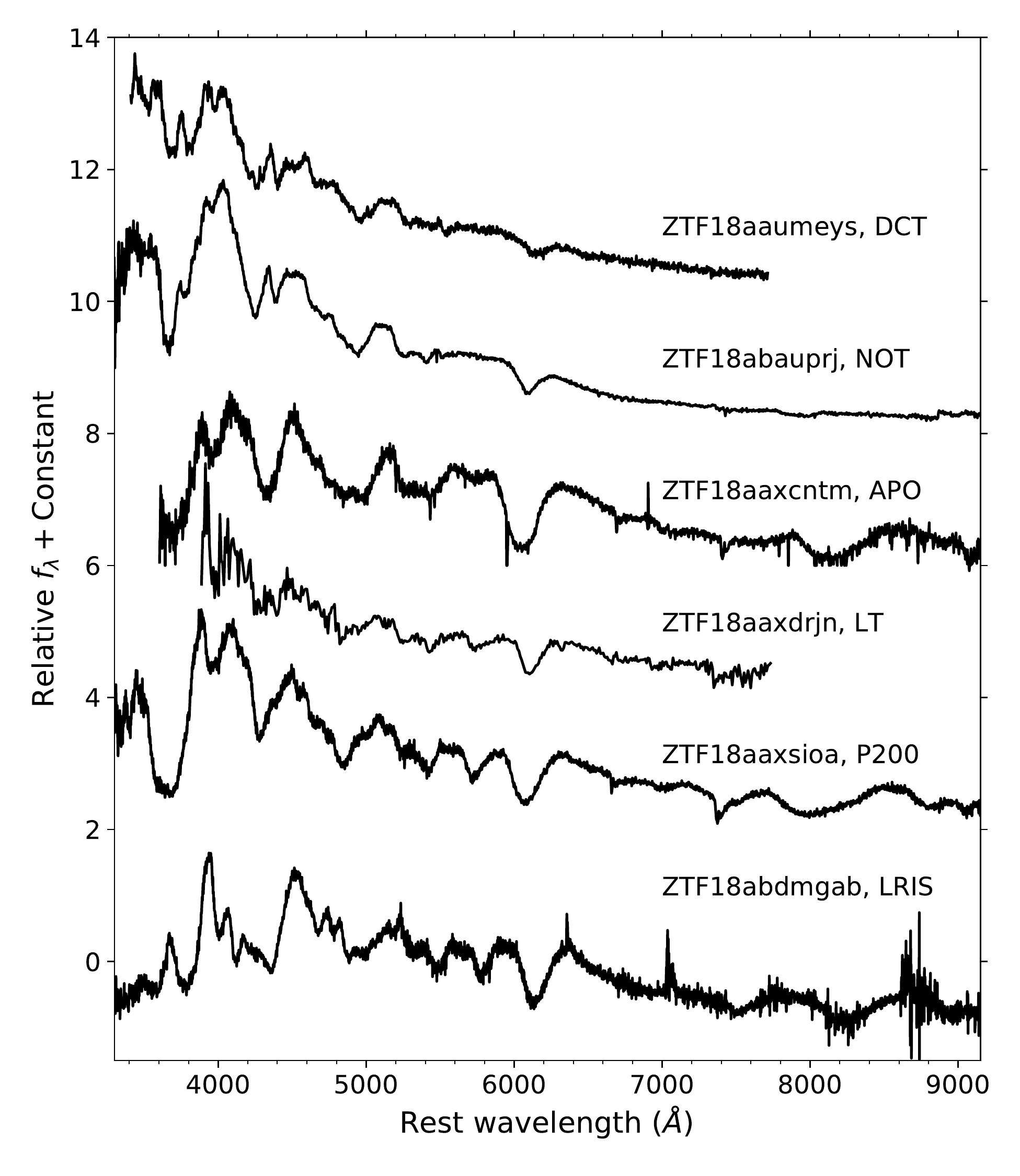}
    \caption{A subset of the spectra used for classification for six objects in our sample. Details of the spectra are given in Table~\ref{tab:spec}. \label{fig:showspec}}
\end{figure}

Within our sample, 92 SNe are classified solely with SEDM spectra. These spectra will be described in detail in Rigault et al.\ (in prep).\footnote{Most of the SEDM spectra are publically available via TNS.} Below we describe our method for SN subtype classification. Table \ref{tab:spec} provides information on 37 non-SEDM spectra of 34 targets, for which the subtype determination was not solely dependent on SEDM.\footnote{Upon publication, these 37 spectra will be available in electronic format on the Weizmann Interactive Supernova Data Repository (WISEReP, \citealt{Yaron2012}).}  A subset of these spectra are shown in Figure~\ref{fig:showspec}.

\subsubsection{Classification based on SNID}  \label{subsubsec:SNIDclassify}
We use the SuperNova IDentification tool (\texttt{SNID}; \citealt{Blondin2007}) to aid the determination of spectral subtype. \texttt{SNID} determines subtype via template-matching by cross-validation. It divides SNe Ia into five subtypes: Ia-norm, Ia-91T, Ia-91bg, Ia-csm, and Ia-pec. The template bank we used was from the original \texttt{SNID} templates-2.0 set, expanded with spectra of tidal disruption events (TDEs) and superlumionous supernovae (SLSNe). An initial prior on template redshift was given if the host galaxy has an entry in NED. Otherwise we adopted the redshift determined by \texttt{SNID} as the SN redshift, which we report with less confidence in Table \ref{tab:info}. If the top three best matched spectra returned by \texttt{SNID} are of the same subtype, then we consider this subtype as a \textit{reliable} classification. 

\begin{figure*}[ht!]
\includegraphics[width=\textwidth]{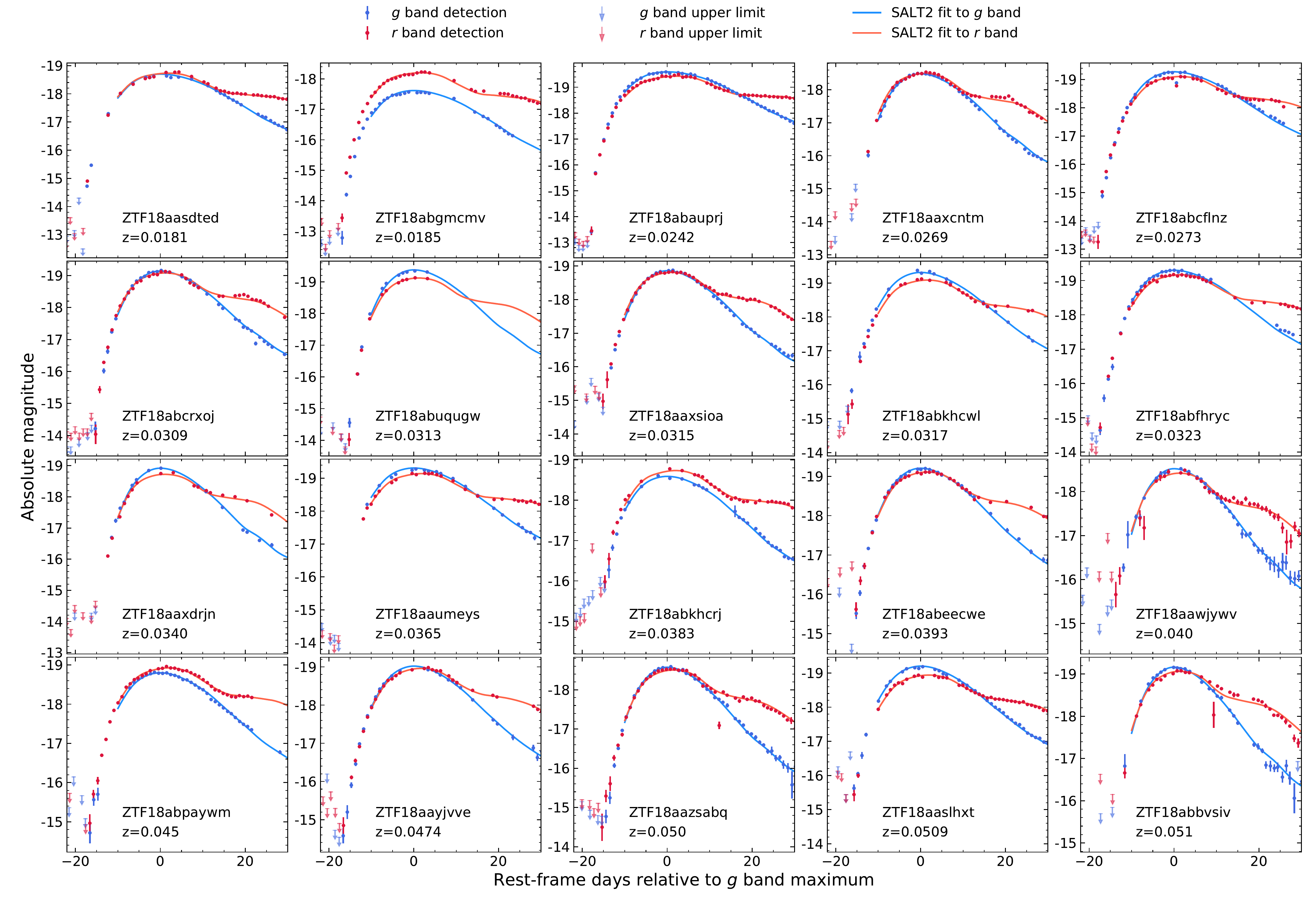}
\includegraphics[width=\textwidth]{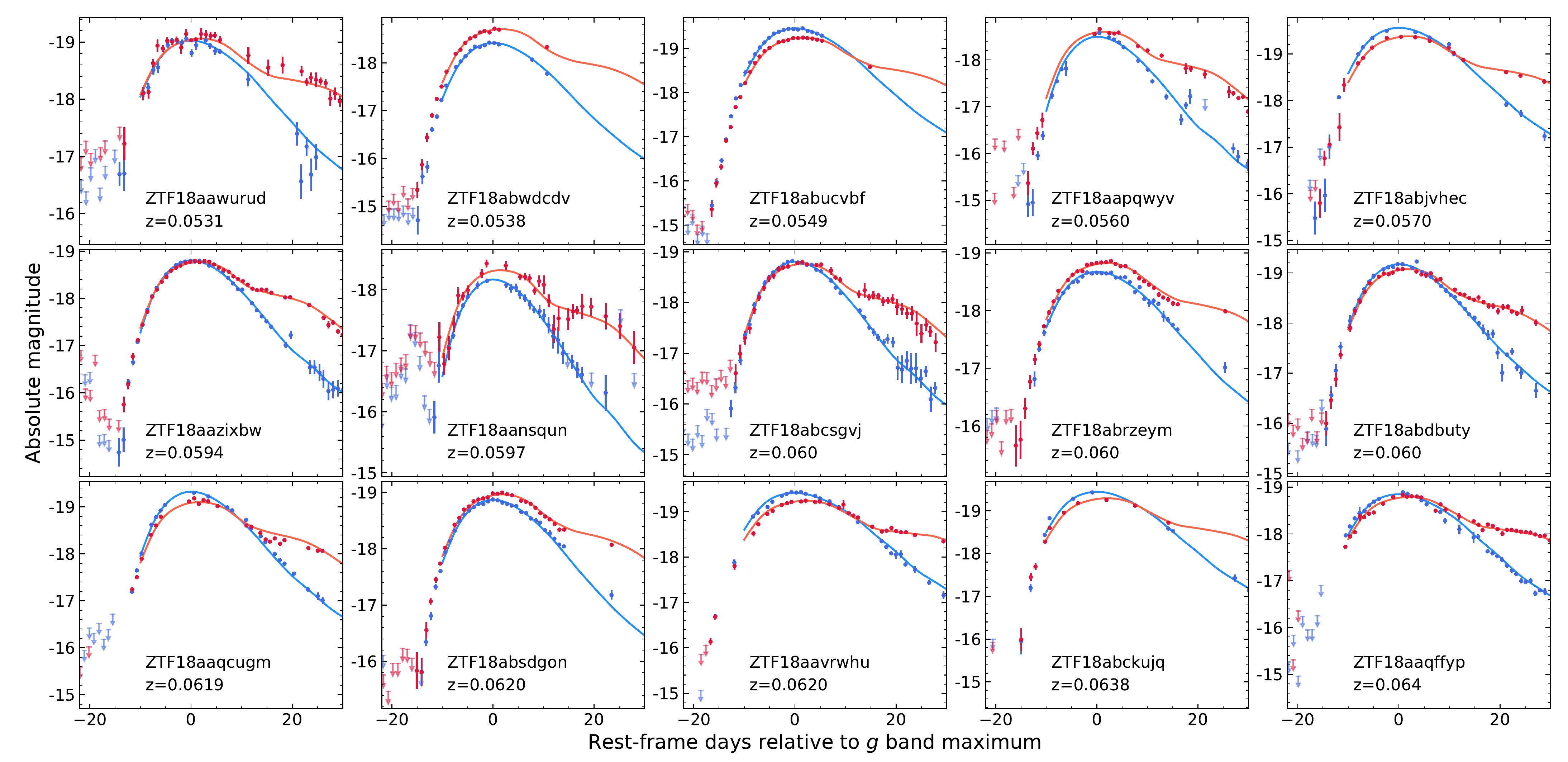}
\caption{P48 light curves of 121 SNe Ia classified as normal($\ast$), 91T-like($\ast$), 99aa-like($\ast$), and 86G-like in our sample. Blue and crimson data points show detections in the $g$ and $r$ bands, respectively. 5$\sigma$ upper limits are shown with downward-pointing arrows. Solid lines are \texttt{SALT2} fits to the data. Absolute magnitude has been corrected for Galactic extinction (host extinction is assumed to be zero for all targets). Single night observations in the same filter are binned (by taking the inverse variance weighted average) for illustration. Targets are ordered by redshift.
\label{fig:lclib}}
\end{figure*}

\begin{figure*}[htbp]
\ContinuedFloat
    \includegraphics[width=\textwidth]{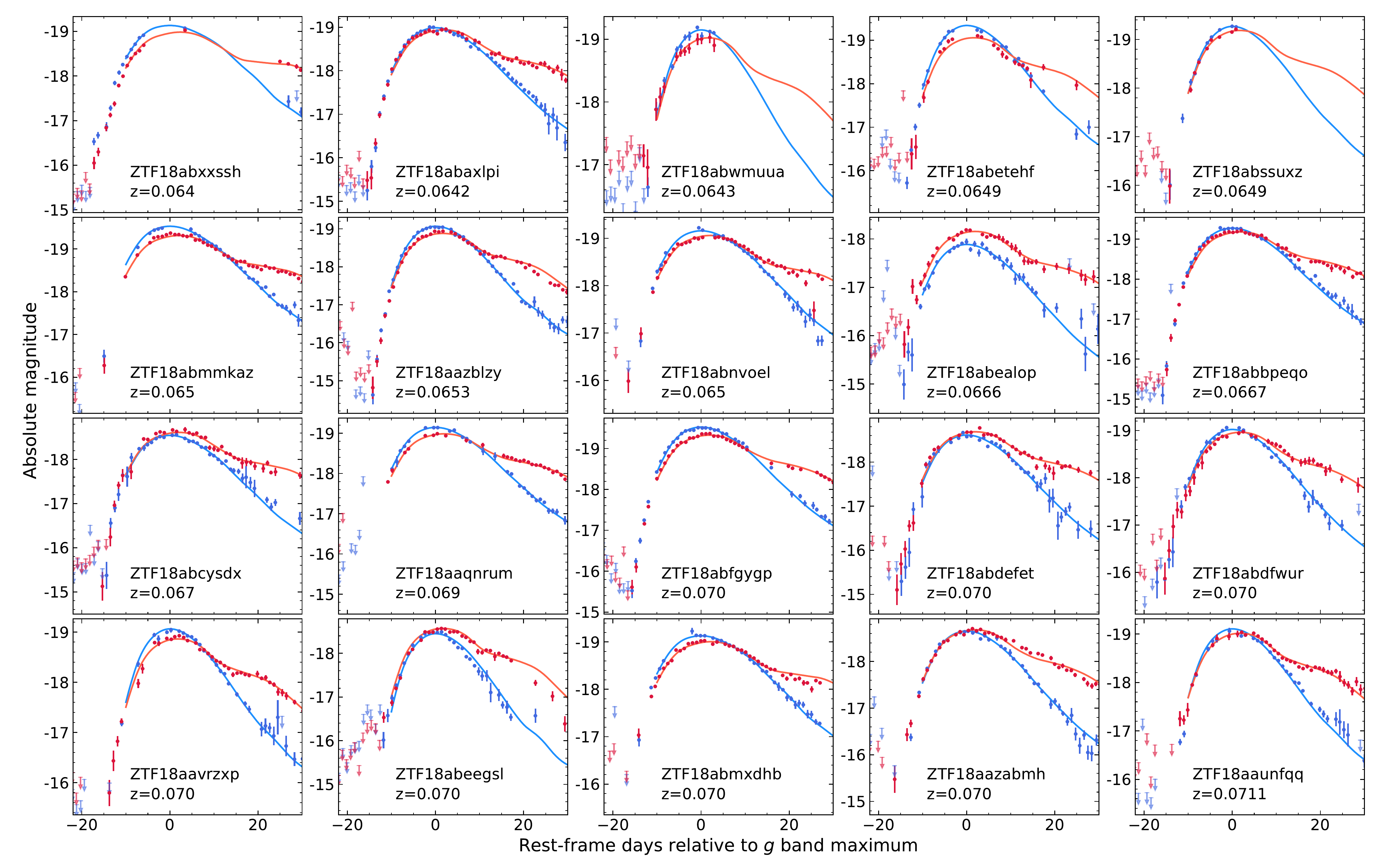}
    \includegraphics[width=\textwidth]{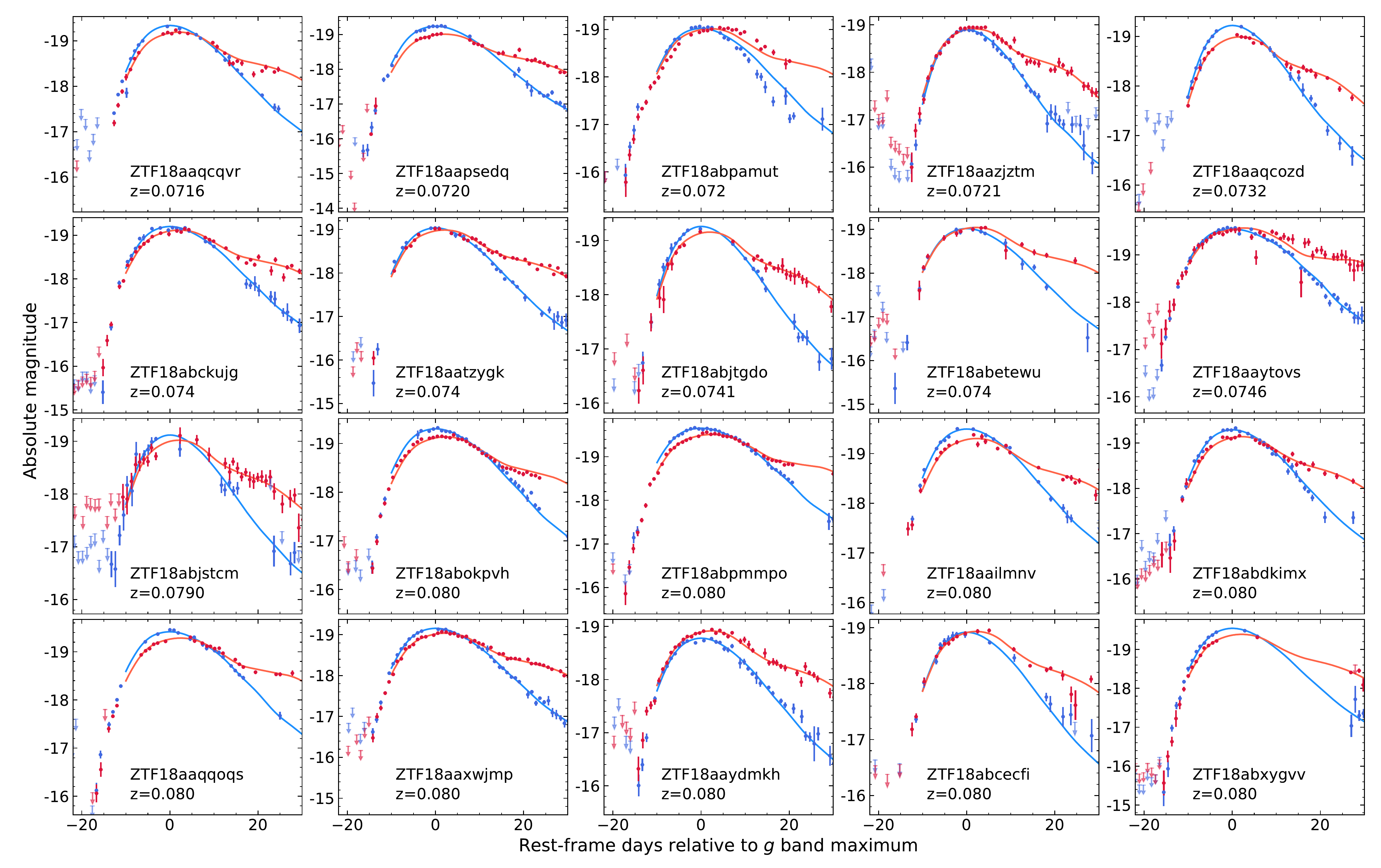}
    \caption{Continued.}
\end{figure*}

\begin{figure*}[htbp!]
\ContinuedFloat
    \includegraphics[width=\textwidth]{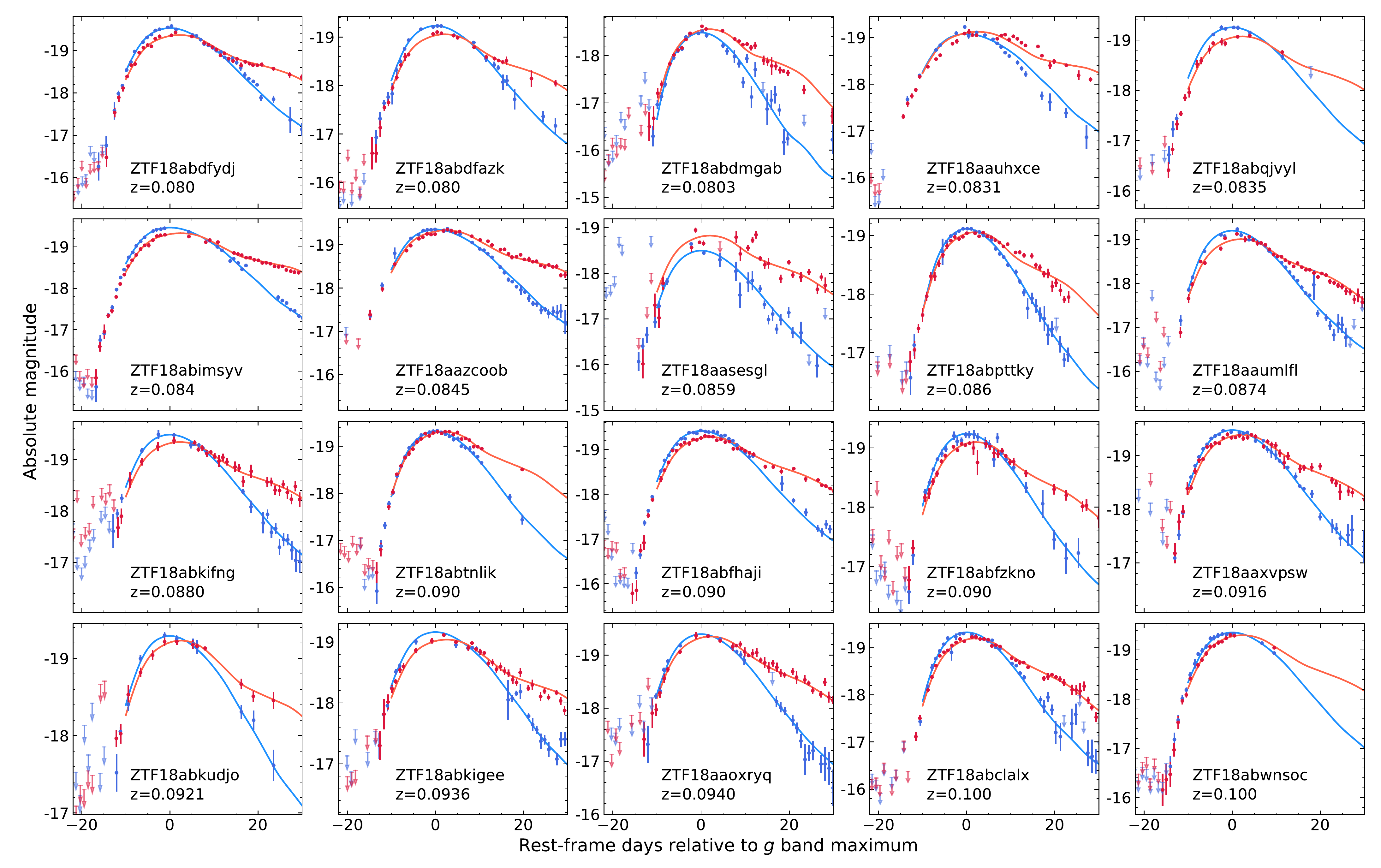}
    \includegraphics[width=\textwidth]{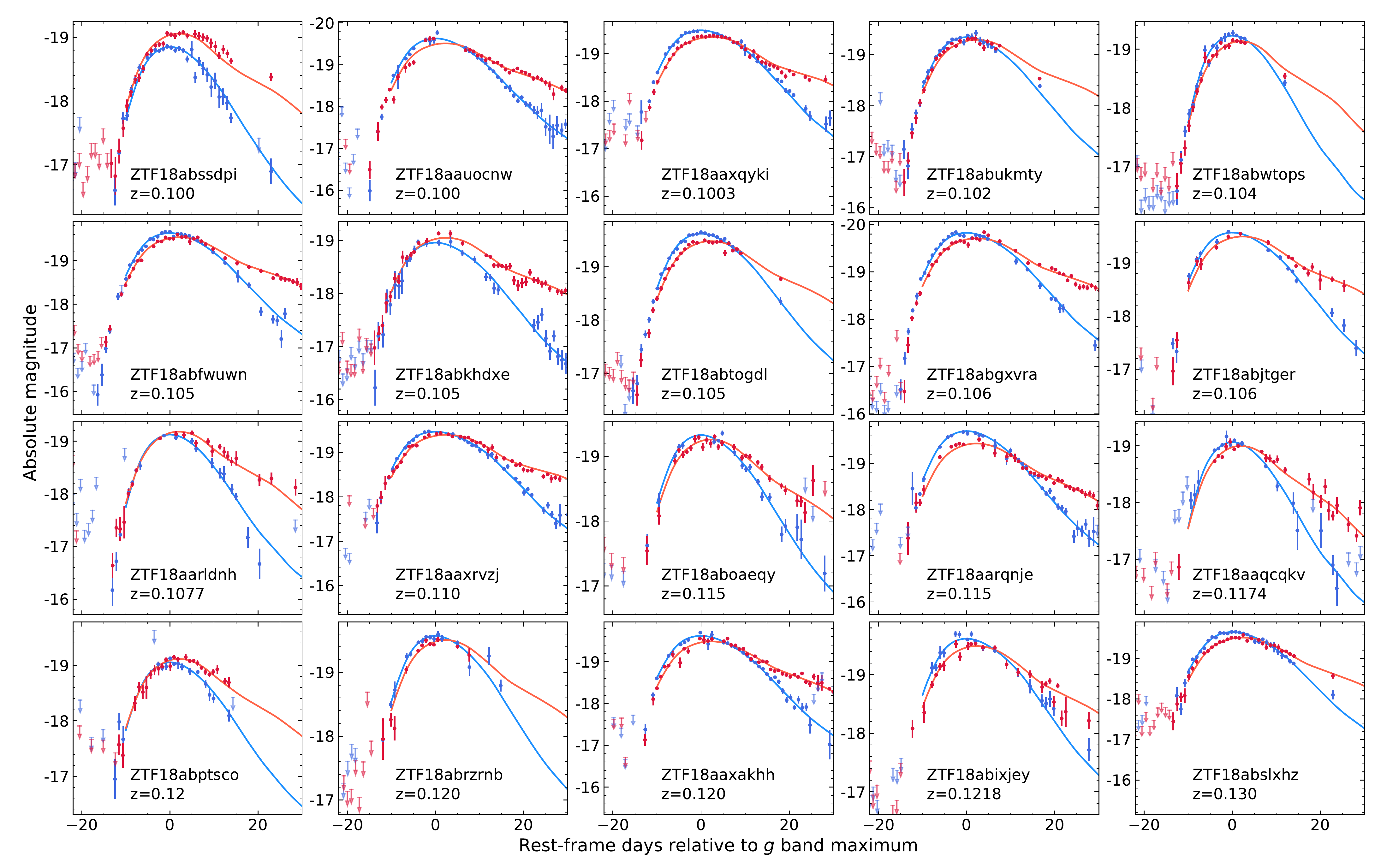}
    \caption{Continued.}
\end{figure*}

\begin{figure}[htbp!]
\ContinuedFloat
    \includegraphics[width=\columnwidth]{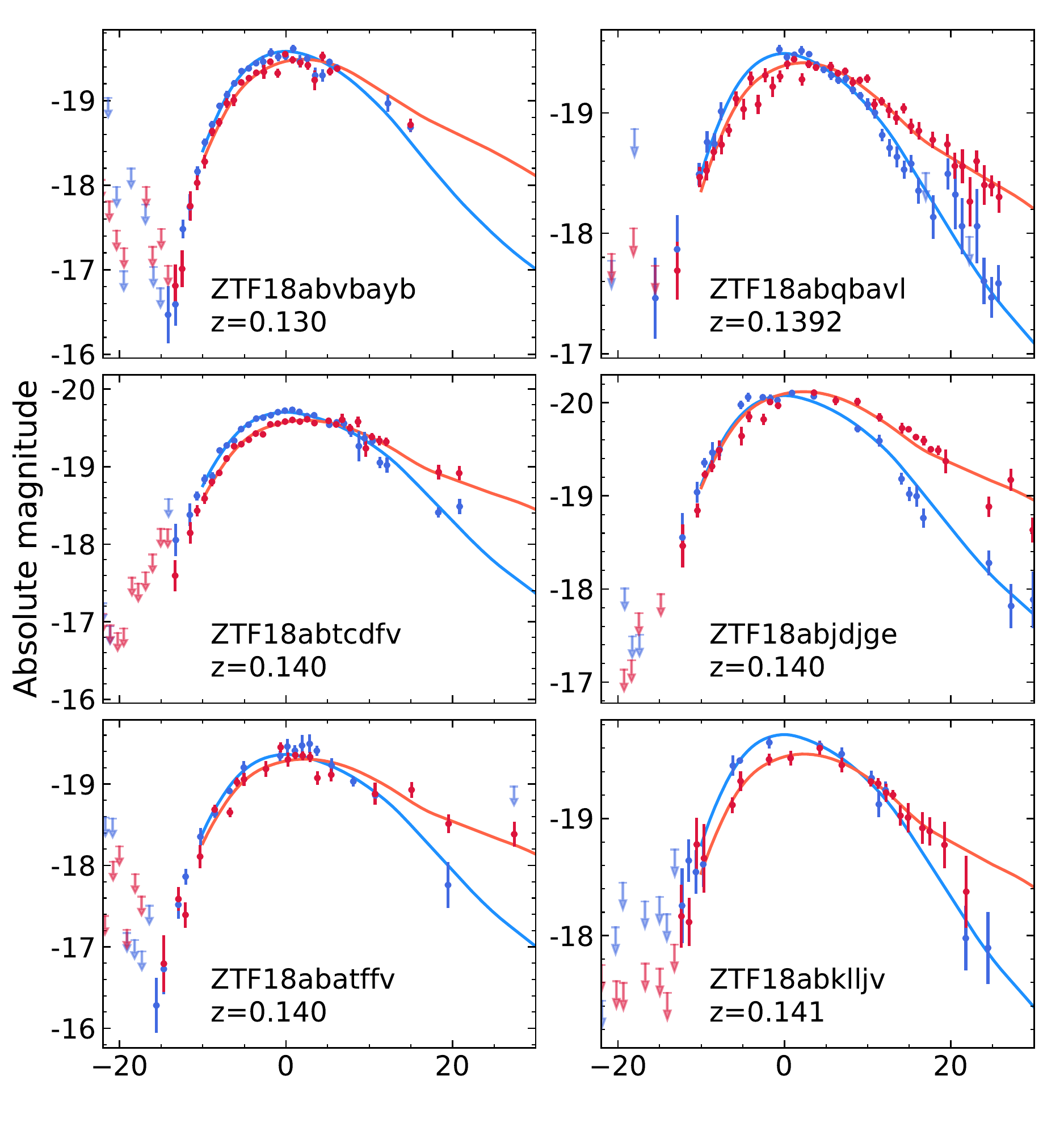}
    \caption{Continued.}
\end{figure}

If the top three best matches returned by \texttt{SNID} were of different subtypes, we adopted the match where the inferred spectral phase (relative to maximum) was closest to the actual phase. If all matches were at similar phases, we further checked the maximum-light absolute magnitude. In cases where $M_{g, \rm max} \gtrsim -19.6$, we \textit{tentatively} classified them as normal SNe Ia (denoted by ``normal$\ast$''), and identified those with $M_{g, \rm max} \lesssim -19.6$ as \textit{potentially} 91T-like events (denoted by ``91T-like$\ast$''). The estimation of absolute magnitude at maximum light can be found in Section \ref{subsec:lc_fitting}. In Table \ref{tab:info}, uncertain subtype determinations are indicated with an asterisk. 

If the classification spectra were obtained sufficiently past maximum (spectral phase $>$15\,d), we chose to put an asterisk at the end of their \texttt{SNID} classification, since a single late time spectrum cannot rule out other subtypes. Given that we do not have a spectrum for ZTF18abptsco, we also considered it to be ``normal$\ast$''. As a result, 85 objects were identified as ``normal'', 26 as ``normal$\ast$'', 8 as the ``91T-like'' subtype, 6 as ``91T-like$\ast$'', 1 as ``91bg-like'', and 1 as ``Ia-CSM''. 

\subsubsection{Subtype Modifications and Reliability}\label{subsubsec:snid_reliability}
\paragraph{99aa-like and 91T-like events}
SN\,1999aa-like events were included in the Ia-91T \texttt{SNID} templates. They are transitional objects between 91T-like and normal SNe Ia, and may represent a spectroscopically distinct subclass \citep{Silverman2012}. Therefore, for targets classified as 91T-like events by \texttt{SNID}, we further inspected their spectra and best matched templates. 99aa-like events were identified via their strong \ion{Ca}{II} H \& K lines. Seven objects were classified as 99aa-like after first being 91T-like events, while one object was classified as 99aa-like$\ast$ after first being 91T-like$\ast$.

\paragraph{91bg-like and 86G-like events} 
The subluminous 86G-like and 02es-like \citep{Ganeshalingam2012} objects were included in the Ia-91bg \texttt{SNID} templates. For the one object (ZTF18abdmgab) classified by \texttt{SNID} as Ia-91bg, we compared its maximum light spectrum with that of SN\,1991bg, SN\,1986G and SN\,2002es. It is clearly more similar to SN\,1986G, so we classify it as such.

\paragraph{Objects with multiple spectra}
Multiple spectra have been obtained for 31 events in our sample. Among them, 17 got more than one spectrum with a \textit{reliable} classification. It is worth checking the consistency of the \texttt{SNID} classifications from different spectra, because it will help answer whether the top three \texttt{SNID} matches having the same subtype means that such a subtype determination is indeed reliable. 


We noticed that among the 17 events, \texttt{SNID} subtypes of 15 objects are consistent with each other ($15=12$ ``normal'' $+ 1$ ``86G-like'' $+ 1$ ``99aa-like''). The two exceptions are ZTF18abauprj and ZTF18abclfee. ZTF18abauprj was classified by a NOT spectrum at $-6$\,d as a ``99aa-like'' event (see Figure \ref{fig:showspec}). This event received an extensive rapid spectroscopic follow-up campaign: a DCT spectrum at $-11$\,d, two APO spectra at $-12$\,d and $-15$\,d, a Keck I spectrum at $-15$\,d, as well as an LT spectrum at $-15$\,d. Although typical features of 99aa-like events (\ion{Fe}{III} multiplets, strong \ion{Ca}{II} H \& K lines, and weak \ion{Si}{II}) clearly exist in the five spectra, the \texttt{SNID} classifications are, however, normal SN Ia. The inconsistency may result from the lack of early 91T-like templates in the \texttt{SNID} database, which suggests that the ``normal'' typing from early-time spectra (phase $<-10$\,d) of seven events\footnote{They are ZTF18aazixbw, ZTF18abcflnz, ZTF18abcrxoj, ZTF18abfhryc, ZTF18abkhcrj, ZTF18abkhcwl, ZTF18abmxdhb.} in Table \ref{tab:info} may be questionable. However, the peak luminosity of these events ($M_{g, \rm max}$ ranges from $-18.5$ to $-19.2$ mag) are  consistent with these being normal SNe. 

The fact that \texttt{SNID} has more normal SN Ia templates leads to a subtype ``attractor'' with the risk for low SNR spectra to be preferentially classified as normal, regardless of their type. When the number of templates per class is highly unbalanced, then it is far more likely that the statistically best match to a low SNR spectrum will occur with the dominant class. Classification biases due to the unbalanced training set was noted by \citet{Blondin2007} and has been demonstrated in subsequent studies (e.g., \citealt{Foley2009, Ostman2011, Silverman2012}).

ZTF18abclfee was classified by an SEDM spectrum at $+2$\,d as a normal SN, but later spectra at $+4$\,d and $+8$\,d were classified as Ia-pec. This is indeed an 02cx-like event and the classification will be justified in Section \ref{subsec:ZTF18abclfee}. We also identify four super-Chandrasekhar (SC) mass explosions in Section \ref{subsec:superChandra}. Although they were misclassified as ``normal($\ast$)'' or ``91T-like$\ast$'' by \texttt{SNID}, they can be distinguished from normal events by their overluminous peak luminosity, lower velocities, and the lack of distinct second maximum (in the red portion of the optical spectrum) typical of normal SNe Ia \citep{Scalzo2010}.

\paragraph{Conclusion}
Among the 127 SNe Ia in our sample, 82 events were identified as ``normal'', 25 as ``normal$\ast$'', 7 as ``99aa-like'', 3 as ``99aa-like$\ast$'', 1 as ``91T-like'',  2 as ``91T-like$\ast$'', 1 as 86G-like, 1 as ``02cx-like'', 1 as ``Ia-CSM'', 2 as ``SC'', and 2 as ``SC$\ast$''. The identification of peculiar events will be presented in Section \ref{sec:peculiar}. The subtype classification given in Column (8) of Table \ref{tab:info} should be relatively reliable, since we have considered both spectroscopic and photometric properties.

\begin{figure}[htbp!]
\includegraphics[width=\columnwidth]{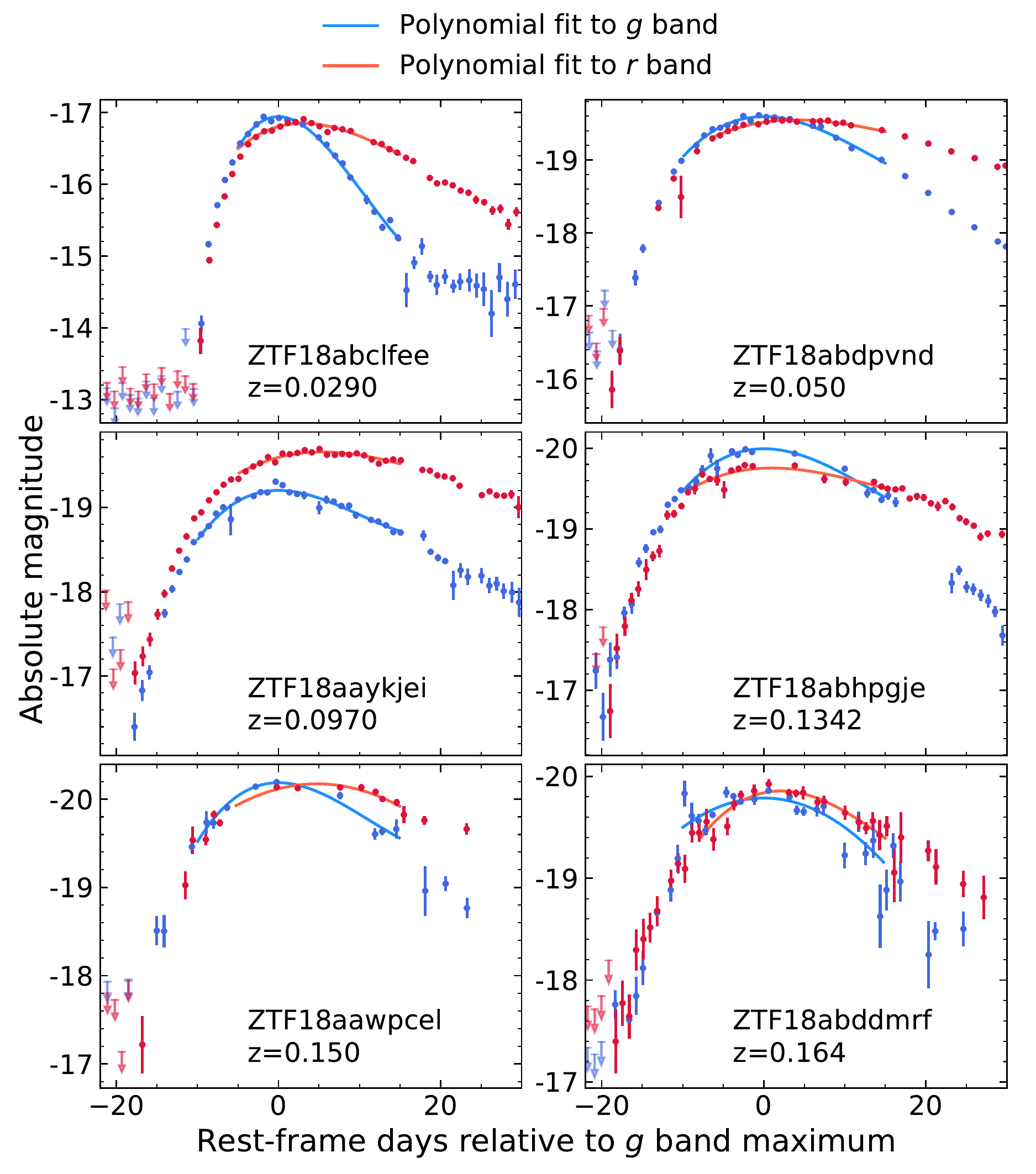}
\caption{P48 light curves of 6 peculiar SNe Ia in our sample. Solid lines are polynomial fits to the data. See Figure \ref{fig:lclib} for symbols. \label{fig:lclib_pec}}
\end{figure}

\subsection{Light Curve Properties} \label{subsec:lc_fitting}
Forced photometry for the 127 SNe Ia in our sample is shown in Figures~\ref{fig:lclib} and~\ref{fig:lclib_pec}. We adopted Eq.\ (\ref{eq:sigma_m}) and Eq.\ (\ref{eq:upplim}) to convert $f_{\rm ratio}$ into the observed magnitude. The absolute magnitude is determined by correcting for the distance modulus and Galactic extinction $E(B-V)$ estimated by \citet{Schlafly2011}, which is build upon \citet{Schlegel1998}. We assume $R_V=3.1$, and integrate the reddening law from \citet{Cardelli1989} over the ZTF filters.

In order to estimate the time of $g$-band maximum ($t_{g, \rm max}$), observed peak magnitude ($m_{g, \rm max}$), and the decline-rate parameter expressed by the decline within 15\,d from maximum in $g$ band ($\Delta m_{15}(g)$), we fit light curves of normal($\ast$), 99aa-like($\ast$), 91T-like($\ast$), and 86G-like objects with \texttt{SALT2} (as shown by the overplotted solid lines in Figure \ref{fig:lclib}). For the 02cx-like, Ia-CSM, and SC($\ast$) SNe we fit the light curves with low-order polynomial functions (as shown by the overplotted solid lines in Figure \ref{fig:lclib_pec}). We convert $m_{g, \rm max}$ into absolute peak magnitude $M_{g, \rm max}$ using the same method as above.

We separate the sample into two groups because \texttt{SALT2} is currently not suitable to satisfactorily determine light curve features of peculiar events. Our choice of low-order polynomial functions follows the fitting technique adopted by \citet{Foley2013} and \citet{Miller2017}. Estimated parameters are reported in Table \ref{tab:phot}.

\subsubsection{Light Curve Fitting with SALT2}
To attempt the \texttt{SALT2} fits, we assumed $R_{V}=3.1$, adopted the Galactic extinction estimate of $E(B-V)$ from \citet{Schlafly2011}, and added a Milky Way dust model (\texttt{CCM89Dust}) \citep{Cardelli1989} to the SN model (\texttt{SALT2Model}) in \texttt{sncosmo} \citep{Barbary2016}. \texttt{SALT2} characterizes the flux density for a given SN as a function of phase $p$ and rest-frame wavelength $\lambda$ as:
\begin{equation}
    f(p, \lambda) = x_0  \left[  \mathcal{M}_0(p, \lambda) +x_1 \mathcal{M}_1(p, \lambda) \right] e^{c \cdot \mathcal{C_{\rm L}(\lambda)}}\label{eq:salt2}
\end{equation}
where $x_0$, $x_1$, and $c$ are the normalization, shape, and color parameters, respectively. The mean spectral sequence $\mathcal{M}_0$, the first order deviation around the mean sequence $\mathcal{M}_1$, and the average color-correction law $\mathcal{C_{\rm L}}$ were trained on photometric and spectroscopic data of known SNe Ia (see \citealt{Guy2007} for details). 

In addition to $t_{g, \rm max}$, $M_{g, \rm max}$, and $\Delta m_{15}(g)$, we also obtained the expected $\Delta m_{15}(B)$ and $M_{B, \rm max}$ from the \texttt{SALT2} fitted parameters for objects shown in Figure \ref{fig:lclib}. 

\subsubsection{Light Curve Fitting with Polynomial Functions}
We interpolated the observed photometry of peculiar events with low-order polynomial fits. The degree of the polynomial used to fit the light curve is chosen between 2 and 3, optimized using the Bayesian information criteria (BIC; \citealt{Schwarz1978}). According to BIC, another parameter should only be added (by increasing the order of the polynomial) if it decreases the $\chi^2$ with at least the natural logarithm of the number of data points. For the five overluminous events (1 Ia-CSM and 4 SC($\ast$)), the time range used in the fit is from $-10$\,d to $+16$\,d (in rest frame) relative to maximum light. For the 02cx-like object we only fit data points from $-5$\,d to $+16$\,d (in rest frame) relative to maximum light.

We obtained the covariance matrix of the polynomial coefficients ($\mathbf{Cov}$) with \texttt{numpy}'s \texttt{polyfit} function. To estimate uncertainties of $t_{g, \rm max}$, $M_{g, \rm max}$, and $\Delta m_{15}(g)$, 100,000 Monte Carlo simulations were run by resampling the polynomial coefficients from the Cholesky decomposition of $\mathbf{Cov}$. 

\subsubsection{The Selection Effect}
\begin{figure}[htbp!]
    \centering
    \includegraphics[width=0.9\columnwidth]{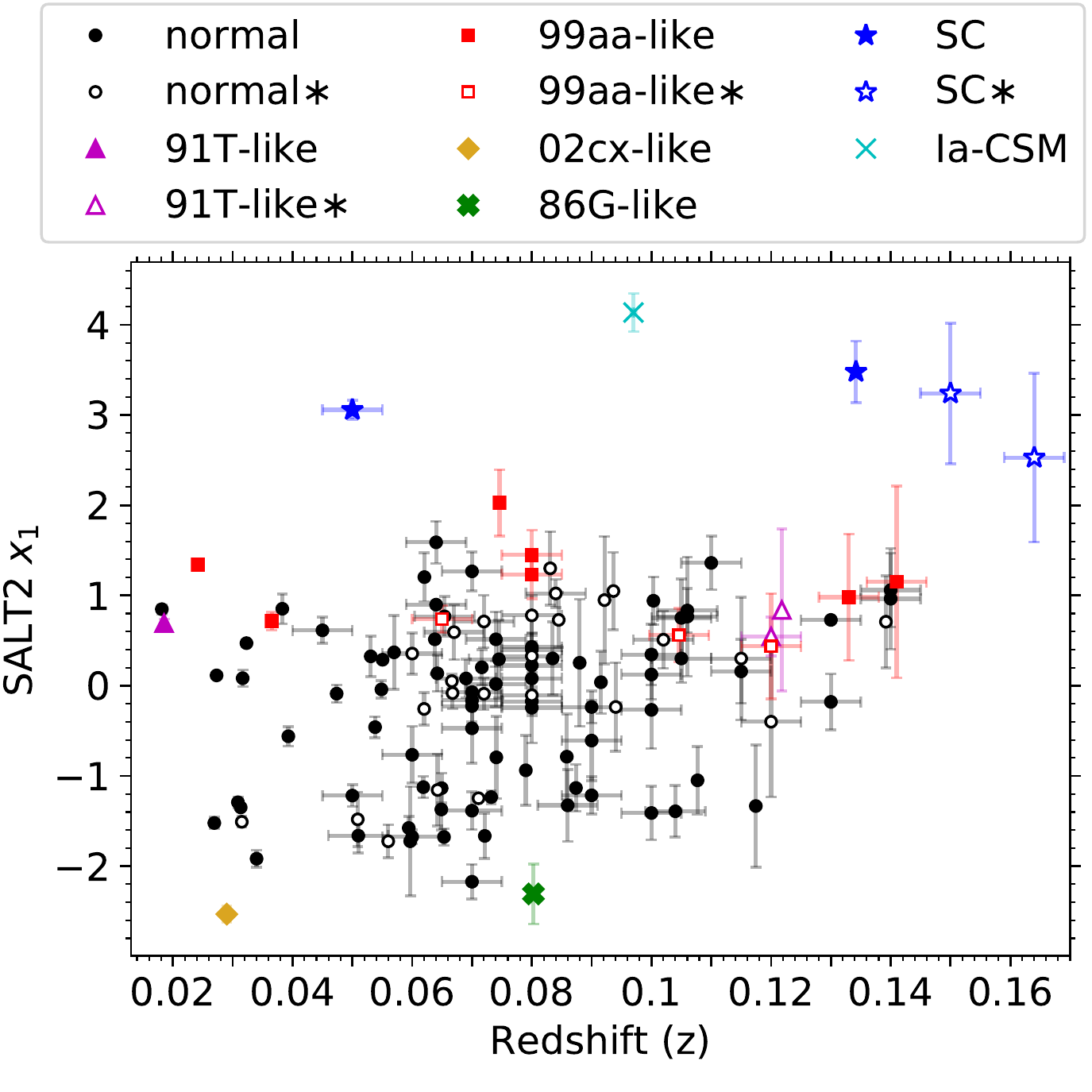}
    \caption{\texttt{SALT2} $x_1$ vs.\ Redshift (z) for the 127 SNe in our sample. The apparent positive correlation between $x_1$ and redshift is due to the Malmquist bias (fainter events, i.e., low $x_1$, are not detected at $z\gtrsim0.1$). Section \ref{sec:peculiar} discusses the peculiar events. \label{fig:x1_c}}
\end{figure}
Figure \ref{fig:x1_c} shows the distribution of 127 SNe on the light curve shape parameter \texttt{SALT2} $x_1$ vs.\ redshift plane. Larger $x_1$ indicates broader light curve width and greater maximum luminosity. We note that the correlation between $x_1$ and $\Delta {m_{15}}(g)$ is sufficiently strong that $x_1$ can be used as a proxy for the light curve decline-rate, and thus peak luminosity (see Figure \ref{fig:x1_dm15}). The median of $x_1$ should be around zero for an unbiased sample. For our sample, $x_1$ is centered at $\sim$0 for 100 SNe with $z\leq0.1$ (median $x_1=0.007$), but shifts to greater values for the 27 objects with $z>0.1$ (median $x_1=0.729$). This is a consequence of Malmquist bias \citep{Malmquist1922} --- at higher redshift, only targets intrinsically more luminous can be detected early enough. The Malmquist bias had been predicted at about the same level from the PTF/iPTF work (see Figure 11 of \citealt{Papadogiannakis2019}).

\subsubsection{The Luminosity-Decline Relation} \label{subsec:phillips}
\begin{figure}[htbp!]
\centering
\includegraphics[width=0.9\columnwidth]{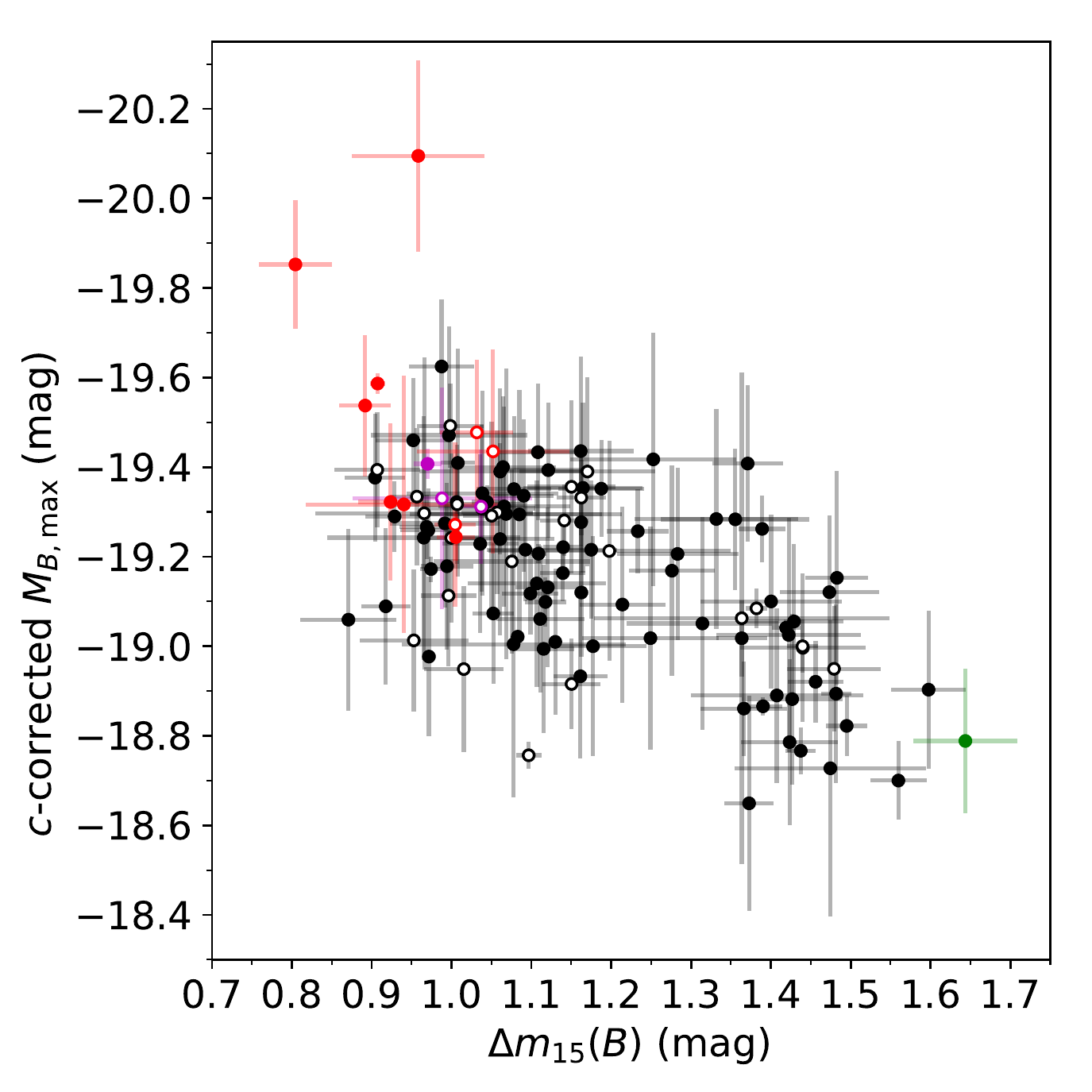}
\caption{Distribution of 121 SNe on the \texttt{SALT2} $c$-corrected $M_{B, \rm max}$ vs. $\Delta m_{15}$ ($B$) plane. Symbol colors follow the same convention as in Figure \ref{fig:x1_c}.  \label{fig:Phillips_Bband} }
\end{figure}

Accurate host extinction estimates are critical for the luminosity-decline relation of SNe Ia \citep{Phillips1993}. An empirical way to get the host extinction corrected absolute $B$-band magnitude is to subtract 3.1 times the \texttt{SALT2} color parameter ($c$) from the Galactic extinction corrected $M_{B, \rm max}$ \citep{Betoule2014}. The $c$-corrected $M_{B, \rm max}$ is plotted against the light curve decline rate $\Delta m_{15}$ ($B$) for 121 SNe in Figure \ref{fig:Phillips_Bband}. Peculiar events shown in Figure \ref{fig:lclib_pec} are not included since their light curves can not be well-fitted by \texttt{SALT2}. There is a fairly tight correlation between luminosity and decline rate for normal SNe Ia plotted as filled black circles.

\section{Peculiar Events} \label{sec:peculiar}
\begin{figure}[htbp!]
\includegraphics[width=0.9\columnwidth]{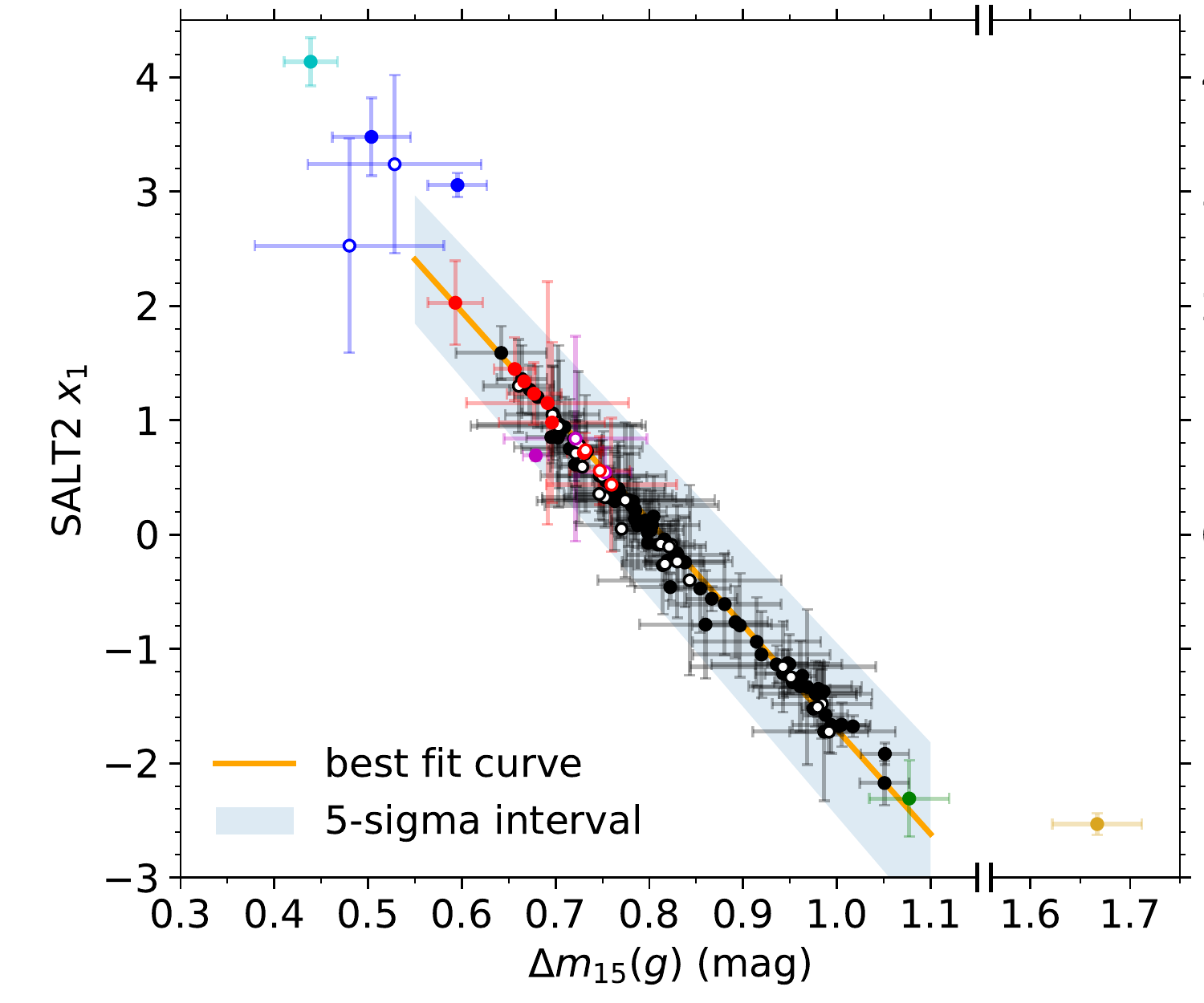}
\caption{Distribution of 127 SNe on the \texttt{SALT2} $x_1$ vs. $\Delta m_{15} (g)$ plane. Symbol colors follow the same convention as in Figure \ref{fig:x1_c}. The solid line shows a linear fit on non-peculiar objects (data shown in black, red, magenta, and green). Note the broken axis.\label{fig:x1_dm15} }
\end{figure}

It has been demonstrated that for peculiar events, in the absence of a constraint on redshift, the fraction of correct subtype matches is low due to the limited number of peculiar templates in the database \citep{Blondin2007}. However, peculiar events can also be selected by their unusual light curve properties. Figure \ref{fig:x1_dm15} shows the distribution of 127 SNe Ia on the \texttt{SALT2} $x_1$ vs. decline rate plane. The light curve shape parameter $x_1$ closely tracks $\Delta m_{15}(g)$ within uncertainties for normal, 99aa-like, 91T-like, and 86G-like SNe in our sample. A linear fit gives 
\begin{equation}
    x_1 = (-9.13\pm 0.09) \times \Delta m_{15} (g) +( 7.43 \pm 0.06).
\end{equation}

There are six events with $x_1>2$. Except for ZTF18aaytovs, which is spectroscopically classified as a 91T-like object (Section \ref{subsec:subtype}), the other five objects are all peculiar overluminous slow decliners, and are discussed below. The one object shown in yellow (ZTF18abclfee) has the smallest value of $x_1$, and declines much faster than other SNe (note the broken $x$ axis). These events can all be classified as ``peculiar,'' below we illustrate their observational characteristics to justify their classification.

\subsection{A Ia-CSM SN: ZTF18aaykjei (SN\,2018crl)} \label{subsec:ZTF18aaykjei}
\begin{figure}[ht!]
    \centering
    \includegraphics[width=\columnwidth]{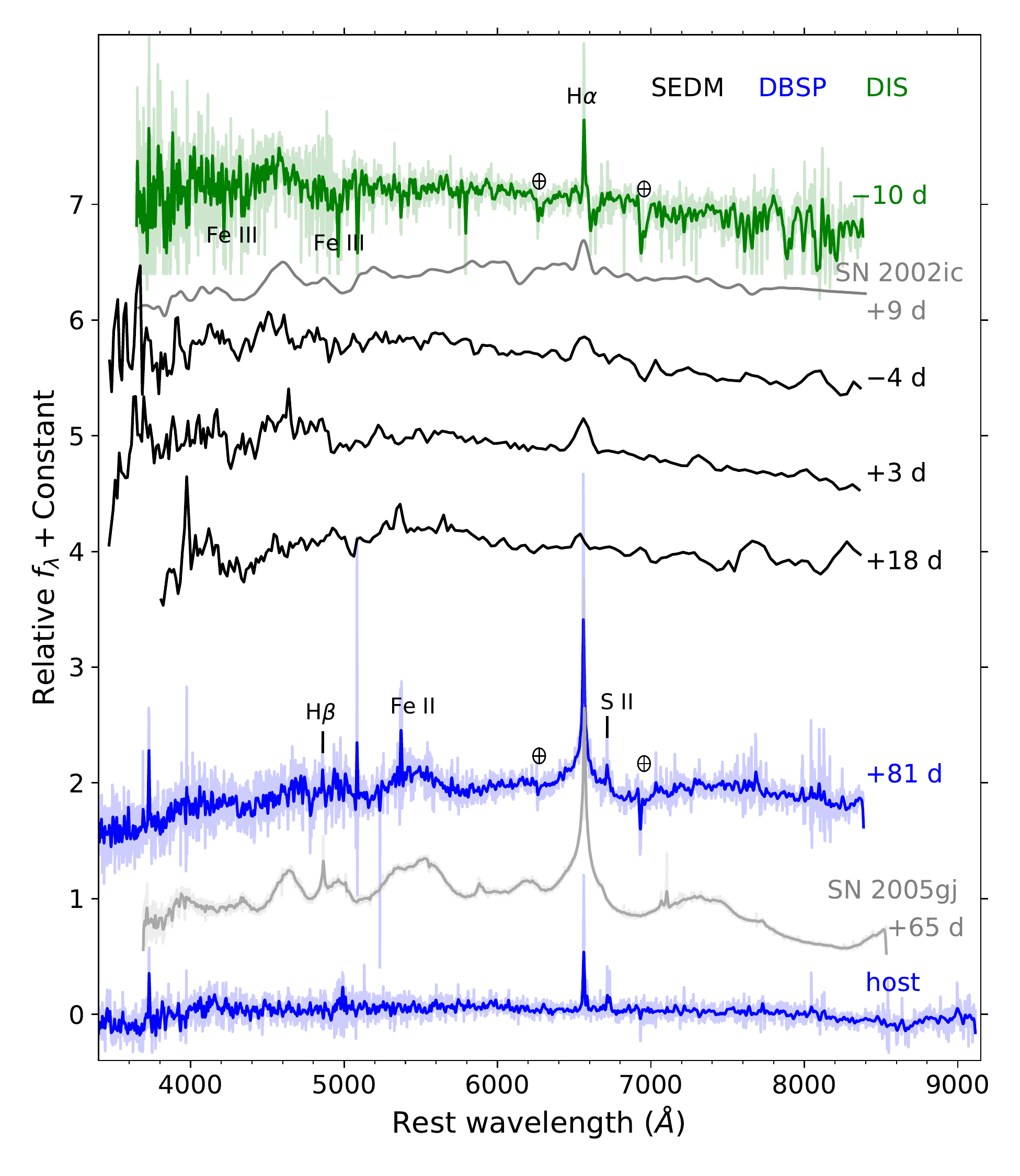}
    \caption{Spectra of ZTF18aaykjei in comparison to two other Ia-CSM objects (SN\,2002ic and SN\,2006gj). The original APO and DBSP spectra are shown in translucent colors, with the overlying solid lines showing the same spectra convolved with an FWHM$=500\,\rm km\,s^{-1}$ Gaussian kernel. The spectrum of its host galaxy is shown at the bottom. 
    \label{fig:ZTF18aaykjei_spec}}
\end{figure}

Ia-CSM is a subclass of SN Ia showing evidence of interaction between the ejecta and the dense circumstellar medium \citep{Dilday2012}. As can be seen in Figure \ref{fig:ZTF18aaykjei_spec}, the spectra of ZTF18aaykjei are dominated by H$\alpha$ emission at all epochs. Its early time APO spectrum matches to the prototype of Ia-CSM SNe (SN\,2002ic), whose spectral features resembled 1991T-like events, but diluted in strength \citep{Hamuy2003}. Absorption and emission lines of intermediate-mass elements (IMEs) and iron-peak elements (IPEs) are present in the blue portion of the two spectra at $-4$\,d and $+3$\,d, but the \ion{Si}{II} lines commonly seen in normal SNe Ia spectra are ``veiled'' by continuum radiation. The noisy $+18$\,d SEDM spectrum only matches to a Sb type galaxy spectrum in the \texttt{SNID} database. Its late-time DBSP spectrum matches to the Ia-CSM object SN\,2005gj \citep{Aldering2006}, where emission lines from overlapping IPEs (mostly \ion{Fe}{II}) are prominent. 

The redshift ($z=0.0970$) of ZTF18aaykjei is measured from the H$\alpha$, [\ion{N}{II}] $\lambda\lambda6548$, 6583, and [\ion{S}{II}] $\lambda\lambda6716$, 6731 (nebular) lines in the spectrum of its host galaxy (SDSS J161938.91$+$491104.7) obtained by DBSP on May 24 2019. At $+81$\,d, the H$\alpha$ emission line profiles have a narrow component on top of a broad (FWHM $\approx 1020\rm \, km \, s^{-1}$) base, much greater than the H$\alpha$ FWHM of the host-only spectrum ($97\rm \, km \, s^{-1}$). The spectra presented here have relatively low resolution, so we do not expect to resolve the P-Cygni profiles seen in some other Ia-CSM SNe.

We show the light curve of ZTF18aaykjei in the middle left panel of Figure \ref{fig:lclib_pec}. It peaked at $-19.19\pm0.04$ mag in $g$ and $-19.65\pm0.04$ mag in $r$ (only Galactic extinction is corrected in these estimates). Its red color even at early times suggests that the host extinction may be non-negligible. The peak luminosity of ZTF18aaykjei is consistent with other objects in the Ia-CSM class \citep[$-21.3$ mag $\leq M_{R}\leq -19$ mag,][]{Silverman2013}.

\subsection{Super-Chandrasekhar Explosions}\label{subsec:superChandra}
There are four events in our sample with $x_1 \approx 3$: ZTF18abhpgje, ZTF18abdpvnd, ZTF18aawpcel, and ZTF18abddmrf. Their light curves are displayed in Figure \ref{fig:lclib_pec}. Among them, the former two events are considered to be super-Chandrasekhar mass explosions (termed as ``SC'' in Table \ref{tab:info}) with evidence from spectroscopy, while the latter two events are classified as candidate super-Chandrasekhar SNe (denoted as ``SC$\ast$'' in Table \ref{tab:info}) based only on their light curves.

\subsubsection{ZTF18abhpgje (SN\,2018eul)}
\begin{figure}[ht!]
    \centering
    \includegraphics[width=\columnwidth]{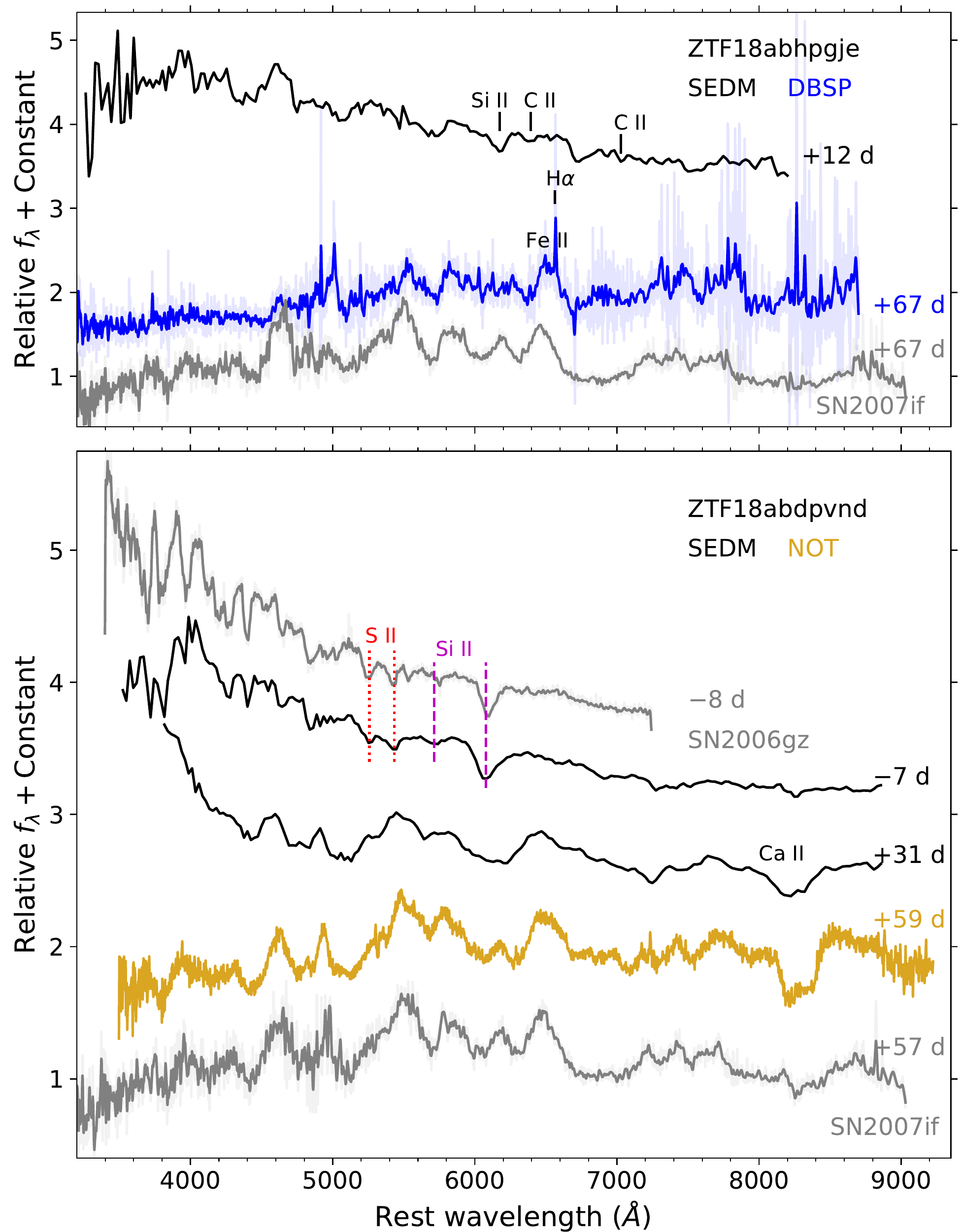}
    \caption{
    Upper panel: Spectra of ZTF18abhpgje. \ion{Si}{II} and \ion{C}{II} lines at $8000\,\rm km\,s^{-1}$ are marked.
    Bottom panel: Spectra of ZTF18abdpvnd. \ion{Si}{II} features at $13,000\,\rm km\,s^{-1}$ and \ion{S}{II} features at $10,000\,\rm km\,s^{-1}$ are marked.
    Spectra of SN\,2009gz and SN\,2007if are obtained from WISEReP and are shown in grey. The DBSP spectrum is smoothed by a Gaussian kernel with FWHM$=500\,\rm km\,s^{-1}$. \label{fig:superChandra_spec}}
\end{figure}

The redshift of ZTF18abhpgje was measured to be $0.134$ from the host H$\alpha$ emission in its DBSP SN spectrum (upper panel of Figure \ref{fig:superChandra_spec}). After correcting for Galactic extinction, it peaked at $-19.99\pm0.04$ mag in $g$ and at $-19.74\pm0.03$ mag in $r$. \ion{Si}{II} $\lambda6355$ and \ion{C}{II} $\lambda \lambda6580$, 7234 absorption features can be identified in its SEDM spectrum, with velocities of $\sim$8000\,$\rm km\,s^{-1}$. The measured velocity is slower than normal SNe Ia ($\sim$10,000\,$\rm km\,s^{-1}$) but similar to some other super-Chandra mass SN candidates: at $+12$ d, SN\,2007if has $v($\ion{Si}{II} $\lambda6355)\sim$8600\,$\rm km\,s^{-1}$ \citep{Scalzo2010}, and SN\,2009dc has $v($\ion{Si}{II} $\lambda6355)\sim$7500\,$\rm km\,s^{-1}$ \citep{Yamanaka2009}. The extreme luminosity and low velocity have been interpreted as the result of either high gravitational binding energy \citep{Howell2006} or the deceleration of the outer layers of ejecta by a massive envelope surrounding the progenitor \citep{Scalzo2010}. The best match of its late time DBSP spectrum is SN\,2017if at $+67$ d (also shown in the upper panel of Figure \ref{fig:superChandra_spec} for comparison), which further supports the argument that ZTF18abhpgje has a super-Chandrasekhar mass progenitor.

\subsubsection{ZTF18abdpvnd (SN\,2018dvf)}
The redshift of ZTF18abdpvnd was inferred to be 0.05 by \texttt{SNID}. After accounting for Galactic extinction, it peaked in $g$ and $r$ band at $-19.59\pm0.26$ mag and $-19.56\pm0.23$ mag, respectively\footnote{Note that the reported uncertainties of peak magnitudes also takes the uncertainty of redshift into consideration, and thus are relatively large.}. This is slightly fainter than other overluminous peculiar objects in our sample, but is still significantly more luminous than normal SNe Ia. As can be seen in the bottom panel of Figure \ref{fig:superChandra_spec}, at $-7$ d, intermediate-mass elements (\ion{Si}{II} $\lambda5872$, 6355 at $\sim 13,000\,\rm km\,s^{-1}$, \ion{S}{II} at $\sim 10,000\,\rm km\,s^{-1}$) are clearly present in its SEDM spectrum. This is similar to the early time velocity of SN\,2006gz \citep{Hicken2007}, which is another well-observed SN with a super-Chandrasekhar mass progenitor. There are other well-matched spectral features between ZTF18abdpvnd and SN\,2006gz. At $+31$ d, although the blue side of ZTF18abdpvnd's SEDM spectrum may be affected by calibration issues, \ion{Ca}{II} can be detected on the red side. The best match of its late time NOT spectrum is SN\,2017if at $+57$ d.

\subsubsection{ZTF18aawpcel (SN\,2018cir) and ZTF18abddmrf (SN\,2018dsx)}
The redshifts of ZTF18aawpcel and ZTF18abddmrf were inferred to be 0.150 and 0.164 by \texttt{SNID}, respectively. We note that at such high redshifts  $\sigma_z \gtrsim 0.01$. The $+0$\,d SEDM spectrum of ZTF18aawpcel shows \ion{Si}{II} at $\sim (14,000\pm3000)\,\rm km\,s^{-1}$, which is consistent with typical peak-time velocity of some normal SNe Ia \citep{Blondin2012}. The $+37$\,d SEDM spectrum of ZTF18abddmrf also matches to normal SNe Ia templates. However, their extreme luminosity (ZTF18aawpcel: $M_{g, \rm max}=-19.95\pm0.11$ mag, $M_{r, \rm max}=-19.91\pm0.11$ mag; ZTF18abddmrf: $M_{g, \rm max}=-19.79\pm0.11$ mag, $M_{r, \rm max}=-19.85\pm0.08$ mag) and slow declining rates make them good candidates for super-Chandrasekhar mass explosions. 
    
\subsection{An 02cx-like event: ZTF18abclfee (SN\,2018cxk)}\label{subsec:ZTF18abclfee}
\begin{figure}[ht!]
    \centering
    \includegraphics[width=\columnwidth]{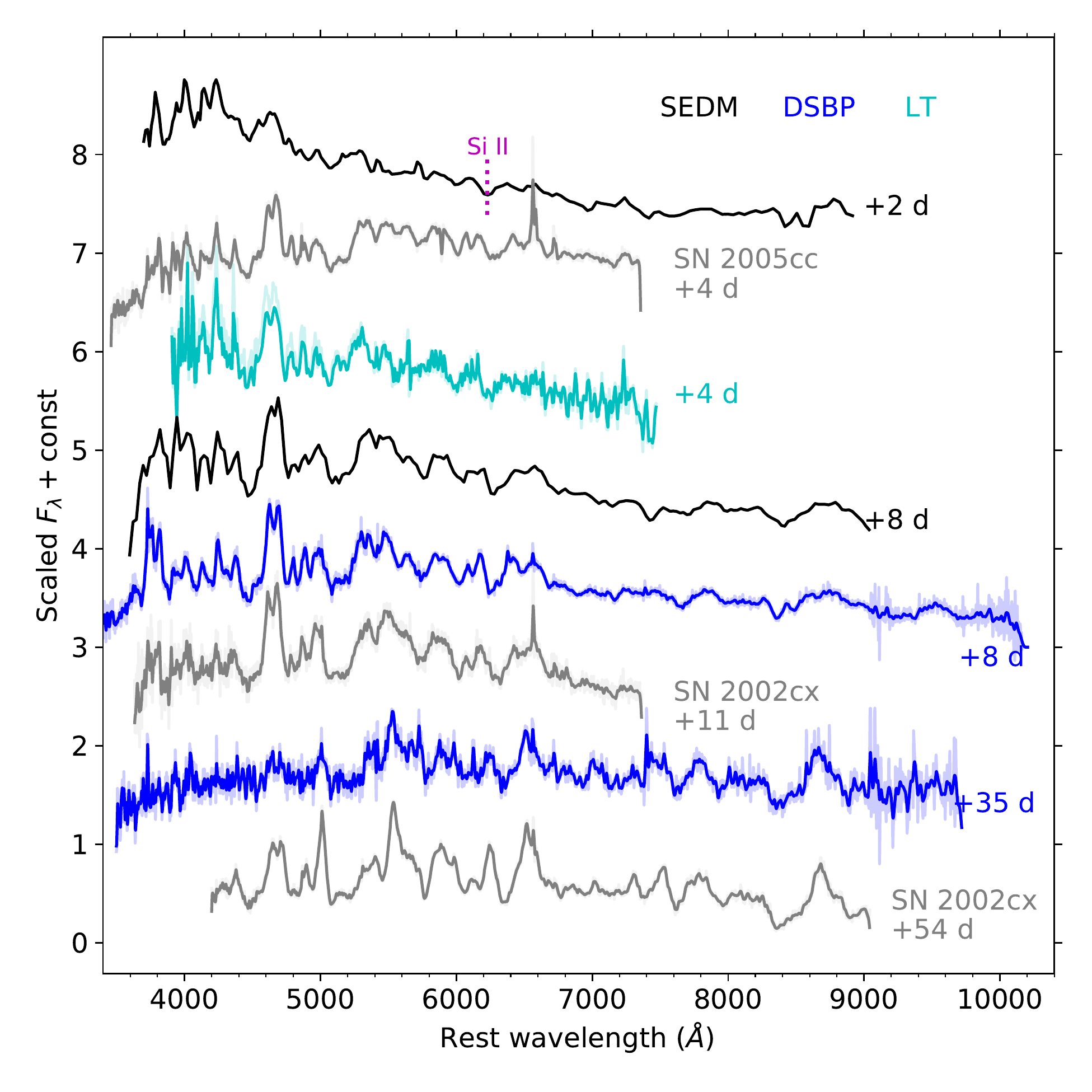}
    \caption{Comparison of spectra of ZTF18abclfee to SN 2002cx and SN 2005cc. The magenta dotted line indicates \ion{Si}{II} $\lambda$6355 at a velocity of 6,000 $\rm km \, s^{-1}$. Non-SEDM spectra are smoothed by a Gaussian filter with with FWHM$=500\,\rm km\,s^{-1}$.\label{fig:ZTF18abclfee_spec}}
\end{figure}
ZTF18abclfee is an 02cx-like event at $z=0.029$ (redshift measured from host H$\alpha$ emission in its $+35$ d DBSP spectrum). This subclass is also termed ``Type Iax Supernovae'' (SN Iax) \citep{Foley2013}. Its photometric and spectroscopic properties are concordant with the criteria of this subclass \citep{Foley2013, White2015}: (1) there is no evidence of hydrogen in any spectra; (2) the \ion{Si}{II} $\lambda$6355 velocity in the $+2$\,d SEDM spectrum is $\sim$6000\,$\rm km \, s^{-1}$, which is much slower than normal SNe Ia; (3) it shows spectral similarity with other 02cx-like events, as can be seen in Figure \ref{fig:ZTF18abclfee_spec}; (4) it is a fast-declining, low-luminosity event.

We do not detect narrow \ion{Na}{I} D at the redshift of 0.029 in any of our spectra, and therefore we assume that host-galaxy extinction is negligible. This assumption is supported by the observed blue color of ZTF18abclfee at peak (upper left panel of Figure \ref{fig:lclib_pec}). After correcting for foreground Galactic extinction, we find that ZTF18abclfee peaked at $M_{g, \rm max}=-16.93\pm0.06$ mag, $M_{r, \rm max}=-16.85\pm0.02$ mag. The decline rates inferred from polynomial fits are $\Delta m_{15}(g) = 1.76\pm 0.08$ mag and $\Delta m_{15}(r) = 0.56\pm 0.02$ mag, which are consistent with other 02cx-like objects (see Figure 8 of \citealt{Miller2017}). The color evolution of ZTF18abclfee is also consistent with the relations shown in figure 9 of \citet{Miller2017}: at maximum light this event has ($g-r$)$_{\rm max}\approx-0.1$ mag, which is bluer than 91bg-like events and similar to normal SN Ia; $\Delta(g-r)_{10} = (g-r)_{+10\rm \,d} - (g-r)_{\rm max} \approx 0.8$ mag, which is similar to 91bg-like events but much greater than $\Delta(g-r)_{10}$ of normal SN Ia.

\section{Summary} \label{sec:summary}
In this paper, we have presented an initial data release for ZTF-discovered SNe Ia with early high-cadence observations. The sample covers 2018 (May--December), and features 127 SNe with dense photometric coverage and early detections in both the $g$ and $r$ bands, allowing an investigation of the initial rise and color evolution. Hence, this sample is well suited for probing the progenitor properties of SNe Ia. 

By comparing our sample with existing samples of low-to-intermediate redshift SNe Ia, we have demonstrated that our sample stands alone in terms of size and early detections. We developed a custom forced-PSF photometry pipeline to extract high quality light curves; these methods can also be applied to other types of extragalactic transients.\footnote{A \texttt{Python} implementation of this method is available at \url{https://github.com/yaoyuhan/ForcePhotZTF}.}

All of the 127 SNe have forced photometry detections earlier than 10\,d prior to $g$-band maximum light (in the rest frame). Their redshifts range from $z=0.01815$ to $z=0.165$, with a median $z=0.074$. The fact that at $z\gtrsim0.1$, the majority of SNe (22/27) have positive values of the light curve shape parameter ($x_1$) suggests that our sample is biased towards overluminous, slow-declining SNe at higher redshift. On average, each SN in our sample has been detected in observations on 46 separate nights.

Although detailed spectroscopic examination is beyond the scope of this paper, we present the spectral sequence of four peculiar events in our sample: one Ia-CSM event ZTF18aaykjei (SN\,2018crl), one 02cx-like event ZTF18abclfee (SN\,2018cxk), two objects with possible super-Chandrasekhar mass progenitors ZTF18abhpgje (AT\,2018eul) and ZTF18abdpvnd (SN\,2018dvf). Futhermore, ZTF18aawpcel (SN\,2018cir) and ZTF18abddmrf (AT\,2018dsx) also exhibit photometric properties that are similar to other super-Chandrasekhar mass explosions. 

\begin{longrotatetable}
\begin{deluxetable*}{lcccccccccccc}
\tabletypesize{\scriptsize}
\tablewidth{0pt} 
\tablecaption{P48 Photometry of 127 SNe Ia.\label{tab:lc}}
\tablehead{
\colhead{ZTF Name}   
& \colhead{JD} 
& \colhead{programid}
& \colhead{fieldid} 
& \colhead{ccdid} 
& \colhead{qid} 
& \colhead{filterid} 
& \colhead{seeing }  
& \colhead{zp} 
& \colhead{$\sigma_{\rm zp}$} 
& \colhead{$f_{\rm mcmc}$} 
& \colhead{$C$} 
& \colhead{$\sqrt{\chi^2_{\nu}}$} \\
\colhead{(ZTF18)}          
& \colhead{}     
& \colhead{}      
& \colhead{}       
& \colhead{}     
& \colhead{}   
& \colhead{}   
& \colhead{(arcsec)}       
& \colhead{(mag)}       
& \colhead{(mag)}  
&  \colhead{(DN)}         
&  \colhead{(DN)}         
&  \colhead{}                    
}
\decimalcolnumbers
\startdata
aazblzy  & 2458291.7770833  &  2  &  722  &  3  &        4  &  2  &  2.164 & 26.185502 & 4.7941 $\times 10^{-6}$        & 1361.0505270 $\pm$ 37.7493381 & $-$0.393 & 1.018 \\
aazblzy  & 2458291.7992593  &  2  &  679  &  15  &       1  &  2  &  2.240 & 26.170234 & 7.9630 $\times 10^{-6}$        & 1367.4679899 $\pm$ 36.4484999 & 7.251 & 1.113 \\
aazblzy  & 2458291.7997338  &  2  &  722  &  3  &        4  &  2  &  2.057 & 26.179769 & 4.9365 $\times 10^{-6}$        & 1377.8058619 $\pm$ 33.4542310 & $-$0.393 & 1.018 \\
aazblzy  & 2458291.8392708  &  2  &  722  &  3  &        4  &  1  &  1.970 & 26.274518 & 8.9916 $\times 10^{-6}$        & 1630.3413509 $\pm$ 35.3272129 & $-$3.349 & 0.904 \\
aazblzy  & 2458292.7180556  &  2  &  722  &  3  &        4  &  1  &  2.103 & 26.299129 & 10.4533 $\times 10^{-6}$       & 1634.6671366 $\pm$ 51.8378991 & $-$3.349 & 0.904 \\
aaqcozd  & 2458257.7669097  &  1  &  716  &  11  &       2  &  2  &  2.565 & 26.275000 & 15.3278 $\times 10^{-6}$       & 1159.2163294 $\pm$ 29.4045185 & $-$0.981 & 0.965 \\
aaqcozd  & 2458257.7678472  &  2  &  716  &  11  &       2  &  2  &  2.554 & 26.275000 & 14.1952 $\times 10^{-6}$       & 1233.6421626 $\pm$ 28.5476163 & $-$0.981 & 0.965 \\
aaqcozd  & 2458257.7778472  &  3  &  716  &  11  &       2  &  2  &  2.695 & 26.275000 & 12.6417 $\times 10^{-6}$       & 1212.3373769 $\pm$ 32.0352867 & $-$0.981 & 0.965 \\
aaqcozd  & 2458257.7787963  &  1  &  716  &  11  &       2  &  2  &  2.643 & 26.275000 & 14.0979 $\times 10^{-6}$       & 1151.8020575 $\pm$ 30.9927667 & $-$0.981 & 0.965 \\
aaqcozd  & 2458257.7797338  &  2  &  716  &  11  &       2  &  2  &  2.541 & 26.275000 & 14.5513 $\times 10^{-6}$       & 1182.5142907 $\pm$ 29.5918902 & $-$0.981 & 0.965 \\
abdpvnd  & 2458364.8118171  &  1  &  646  &  13  &       2  &  2  &  1.916 & 26.106877 & 3.3710 $\times 10^{-6}$        & 1082.7538462 $\pm$ 24.8383520 & $-$999 & $-$999 \\
abdpvnd  & 2458364.8239120  &  2  &  692  &  1  &        4  &  2  &  2.217 & 25.958831 & 2.8078 $\times 10^{-6}$        & 988.7527352 $\pm$ 25.3953545 & $-$999 & $-$999 \\
abdpvnd  & 2458364.8460764  &  2  &  692  &  1  &        4  &  2  &  2.296 & 25.948320 & 3.5620 $\times 10^{-6}$        & 982.4542010 $\pm$ 27.9368325 & $-$999 & $-$999 \\
abdpvnd  & 2458365.8068634  &  2  &  692  &  1  &        4  &  1  &  2.323 & 26.026621 & 6.2705 $\times 10^{-6}$        & 307.9495909 $\pm$ 16.4243875 & $-$999 & $-$999 \\
abdpvnd  & 2458365.8400579  &  2  &  692  &  1  &        4  &  2  &  2.442 & 26.006744 & 2.9185 $\times 10^{-6}$        & 963.6859996 $\pm$ 24.3100430 & $-$999 & $-$999 \\
\enddata
\tablecomments{
Column (3): Program identifier (1 = public survey, 2 = partnership survey, 3 = Caltech survey).
Column (4): fieldid = ZTF field identifier. 
Column (5): ccdid = CCD identifier (from 1 to 16). 
Column (6): qid = Quadrant (CCD-amplifier) identifier (from 1 to 4).
Column (7): filterid = Filter identifier ($1=g_{\rm ZTF}$, $2=r_{\rm ZTF}$). The fcqf ID (Eq. \ref{eq:fcqf}) can be calculated with columns (3)--(7).
Column (8): Photometric seeing (FWHM) at the Palomar Observatory at the time of observation.
Column (9) \& (10): Photometric zero point.
This table is available in its entirety in the machine-readable form.
Column (11): Forced difference image PSF-fit flux. Note that this is the direct measurement from Section \ref{subsec:fmcmc}. Baseline correction described in Section \ref{subsec:baseline} is not applied to these values.
Column (12): history offset in the baseline region.
Column (13): square root of the reduced chi square statistics (Eq. \ref{eq:chi2_red}). 
Note that Column (12) \& (13) are set to $-999$ if $N_{\rm base}<30$. See Section \ref{subsec:baseline} for details.
Only 15 observations of three objects are shown to present format of this table, which is available in its entirety in the machine-readable form.
}
\end{deluxetable*}
\end{longrotatetable}
\begin{deluxetable*}{llcc}
\tabletypesize{\scriptsize}
\tablecaption{Summary of Spectroscopic Observations. \label{tab:spec}}
\tablehead{
\colhead{ZTF Name}   
& \colhead{Telescope} 
& \colhead{UT Date} 
& \colhead{Observers / Reducers}   \\
\colhead{(ZTF18)}              
& \colhead{}     
& \colhead{(YYYY MMDD)}      
& \colhead{}     
}
\startdata
aapqwyv & DCT   & 2018   0506   & C. Ward / T. Hung \\
aaqcozd & NOT   & 2018   0511   & F. Taddia \\
aaqcqkv & P200  & 2018   0517   & A. Ho, Y. Sharma / A. Ho \\
aaqqoqs & P200  & 2018   0517   & A. Ho, Y. Sharma / A. Ho \\
aarldnh & P200  & 2018   0517   & A. Ho, Y. Sharma / A. Ho \\
aarqnje & P200  & 2018   0517   & A. Ho, Y. Sharma / A. Ho \\
aasesgl & P200  & 2018   0517   & A. Ho, Y. Sharma / A. Ho \\
aaslhxt & LT    & 2018   0523   & D. Perley \\
aatzygk & DCT   & 2018   0521   & P. Gatkine / B. Cenko \\
aauhxce & NOT   & 2018   0604   & F. Taddia \\
aaumeys & DCT   & 2018   0521   & P. Gatkine / B. Cenko \\
aaxcntm & APO   & 2018   0613   & M. Graham, D. Bektesevic / M. Graham \\
aaxdrjn & LT    & 2018   0603   & D. Perley \\
aaxqyki & P200  & 2018   0608   & A. Ho, Y. Sharma / A. Ho \\
aaxrvzj & APO   & 2018   0620   & J. Davenport / M. Graham \\
aaxsioa & P200  & 2018   0608   & A. Ho, Y. Sharma / A. Ho \\
aaydmkh & P200  & 2018   0608   & A. Ho, Y. Sharma / A. Ho \\
aaytovs & P200  & 2018   0708   & A. Ho \\
aazabmh & P200  & 2018   0612   & M. Kuhn, Y. Sharma / C. Fremling \\
aazblzy & P200  & 2018   0612   & M. Kuhn, Y. Sharma / C. Fremling \\
aazixbw & NOT   & 2018   0611   & F. Taddia \\
abatffv & Keck1         & 2018   0617   & N. Blagorodnova, K. Burdge, K. De, S. Kulkarni / K. De \\
abauprj & NOT   & 2018   0626   & F. Taddia \\
abdmgab & Keck1         & 2018   0713   & C. Fremling, H. Ko / C. Fremling \\
abfhryc & P200  & 2018   0708   & A. Ho \\
abgmcmv & Keck1         & 2018   0713   & C. Fremling, H. Ko / C. Fremling \\
abixjey & P200  & 2018   0913   & S. Adams, M. Hankins, I. Andreoni / K. De \\
abkifng & P200  & 2018   0821   & C. Fremling, A. Dugas / C. Fremling \\
abqjvyl & P200  & 2018   0912   & A. Dugas / C. Fremling \\
aaykjei & APO   & 2018   0613   & M. Graham, D. Bektesevicc / M. Graham \\
aaykjei & P200  & 2018   0912   & A. Dugas / C. Fremling \\
abclfee & LT    & 2018   0704   & D. Perley / K. Taggart \\
abclfee & P200  & 2018   0708   & A. Ho \\
abclfee & P200  & 2018   0804   & K. De \\
abddmrf & P200  & 2018   0813   & L. Yan / Z. Zhuang \\
abdpvnd & NOT   & 2018   0917   & F. Taddia \\
abhpgje & P200  & 2018   1010   & A. Dugas / A. Ho, Y. Yao \\
\enddata
\tablecomments{Only non-SEDM spectra are included in this table. SEDM spectra will be described in details by one of us (Rigault et al, in prep).}
\end{deluxetable*}

\acknowledgments
Yuhan Yao thanks Yuping Huang for useful discussions on implementing forced photometry, and the Heising-Simons Foundation for financial support. We gratefully thank Yashvi Sharma, Zhuyun Zhuang, Dino Bekte\v sevi\'c, Charlotte Ward, Scott Adams, and Igor Andreoni for help with observations. The authors acknowledge Michael Fausnaugh and Sem\'eli Papadogiannakis for sharing data of the TESS and PTF/iPTF sample, respectively, and an anonymous referee for useful comments that improved the paper. 

This work was supported by the GROWTH project funded by the National Science Foundation under PIRE Grant No.\,1545949. The data presented here were obtained [in part] with ALFOSC, which is provided by the Instituto de Astrofisica de Andalucia (IAA) under a joint agreement with the University of Copenhagen and NOTSA. MB acknowledges support from the Swedish Research Council (Vetenskapsr\aa det), the Swedish National Space Board and the research environment grant ``Gravitational Radiation and Electromagnetic Astrophysical Transients (GREAT)''. A.A.M. is funded by the Large Synoptic Survey Telescope Corporation, the Brinson Foundation, and the Moore Foundation in support of the LSSTC Data Science Fellowship Program, he also receives support as a CIERA Fellow by the CIERA Postdoctoral Fellowship Program (Center for Interdisciplinary Exploration and Research in Astrophysics, Northwestern University). MR has received funding from the European Research Council (ERC) under the European Union's Horizon 2020 research and innovation programme (grant agreement No 759194 - USNAC). A.Y.Q.H. is supported by a National Science Foundation Graduate Research Fellowship under Grant No.\,DGE‐1144469. MLG acknowledges support from the DIRAC Institute in the Department of Astronomy at the University of Washington. The DIRAC Institute is supported through generous gifts from the Charles and Lisa Simonyi Fund for Arts and Sciences, and the Washington Research Foundation. These results made use of the Discovery Channel Telescope at Lowell Observatory. Lowell is a private, non-profit institution dedicated to astrophysical research and public appreciation of astronomy and operates the DCT in partnership with Boston University, the University of Maryland, the University of Toledo, Northern Arizona University and Yale University.  The upgrade of the DeVeny optical spectrograph has been funded by a generous grant from John and Ginger Giovale and by a grant from the Mt. Cuba Astronomical Foundation. MMT Observatory access was supported by Northwestern University and the Center for Interdisciplinary Exploration and Research in Astrophysics (CIERA).

This work is based on observations obtained with the Samuel Oschin Telescope 48-inch and the 60-inch Telescope at the Palomar Observatory as part of the Zwicky Transient Facility project. ZTF is supported by the National Science Foundation under Grant No. AST-1440341 and a collaboration including Caltech, IPAC, the Weizmann Institute for Science, the Oskar Klein Center at Stockholm University, the University of Maryland, the University of Washington, Deutsches Elektronen-Synchrotron and Humboldt University, Los Alamos National Laboratories, the TANGO Consortium of Taiwan, the University of Wisconsin at Milwaukee, and Lawrence Berkeley National Laboratories. Operations are conducted by COO, IPAC, and UW. SED Machine is based upon work supported by the National Science Foundation under Grant No. 1106171.



\bibliography{mybib}
\bibliographystyle{aasjournal}

\end{document}